\documentclass[journal]{IEEEtran}
\usepackage{cite}
\ifCLASSINFOpdf
\else
  \usepackage[dvips]{graphicx}
  \graphicspath{{./eps_all/}}
  \DeclareGraphicsExtensions{.eps}
\fi
\usepackage{tikz}

\usepackage[cmex10]{amsmath}
\usepackage{amsfonts}
\usepackage{cases}

\usepackage{eqparbox}
\usepackage[font=footnotesize, caption=false]{subfig}
\usepackage{fixltx2e}

\ifCLASSOPTIONcaptionsoff
 \usepackage[nomarkers]{endfloat}
 \let\MYoriglatexcaption\caption
 \renewcommand{\caption}[2][\relax]{\MYoriglatexcaption[#2]{#2}}
\fi
 \let\MYorigsubfloat\subfloat
 \renewcommand{\subfloat}[2][\relax]{\MYorigsubfloat[]{#2}}
\usepackage{url}
\usepackage{booktabs}

\def\Z{\hphantom{1}}

\begin{document}
\title{Validation of Geant4-based Radioactive Decay Simulation}
% note positions of commas and nonbreaking spaces (~ ) LaTeX will not break
% a structure at a~ so this keeps an author's name from being broken across

\author{Steffen~Hauf,
    Markus~Kuster,
 Matej~Bati\v{c},
 Zane~W.~Bell,
 Dieter~H.H.~Hoffmann,
 Philipp~M.~Lang,
 Stephan~Neff,
 Maria~Grazia~Pia,
 Georg~Weidenspointner
 and Andreas~Zoglauer% <-this % stops a space
\thanks{Manuscript submitted January 28, 2012. This work has been supported by Deutsches Zentrum f\"{u}r Luft- und Raumfahrt e.V. (DLR) under grants 50 QR 0902 and 50 QR 1102.}% <-this % stops a space
\thanks{S. Hauf and M. Kuster are with European XFEL GmbH, Hamburg, Germany
(e-mail: steffen.hauf@xfel.eu)}% <-this % stops a space
\thanks{D.H.H. Hoffmann, P.––M. Lang and S. Neff are with Institute for Nuclear Sciences, TU Darmstadt, Darmstadt, Germany}% <-this % stops a space
\thanks{M.G. Pia and M. Bati\v{c} are with the INFN Genova, Genova, Italy}% <-this % stops a space
\thanks{G. Weidenspointner is with the Max-Planck Halbleiter Labor, Munich, Germany and the Max-Planck Institut f{\"u}r extraterrestrische Physik, Garching, Germany}% <-this % stops a space
\thanks{A. Zoglauer is with the Space Science Laboratory, University of California, Berkeley, CA, USA}%
\thanks{Z.W. Bell is with the Oak Ridge National Laboratory, Oak Ridge, TN, USA}}%

\maketitle

\begin{abstract}
%\boldmath
Radioactive decays are of concern in a wide variety of applications using Monte-Carlo simulations. In order to properly estimate the quality of such simulations, knowledge of the accuracy of the decay simulation is required. We present a validation of the original Geant4 Radioactive Decay Module, which uses a per-decay sampling approach, and of an extended package for Geant4-based simulation of radioactive decays, which, in addition to being able to use a refactored per-decay sampling, is capable of using a statistical sampling approach. The validation is based on measurements of calibration isotope sources using a high purity Germanium (HPGe) detector; no calibration of the simulation is performed. For the considered validation experiment equivalent simulation accuracy can be achieved with per-decay and statistical sampling.
\end{abstract}


% \fi
%\IEEEpeerreviewmaketitle
\IEEEpeerreviewmaketitle

\section{Introduction}

Radioactive decays and the resulting radiation play an important role for many experiments, either as an observable, as a background source, or as a radiation hazard. Detailed knowledge of the radiation in and around an experiment and its detectors is thus required for successful measurements and the safety of the experimentalist. At the same time the increasing complexity of experiments often makes it prohibitively expensive, if not impossible, to completely determine an experiment's radiation characteristics and response from measurements alone. This is especially true during the design phase of a new project --- much of the hardware may either not be available or the environmental conditions the experiment will be subjected to cannot be replicated (e.g. in space-based observatories). In order to circumvent these limitations it has become increasingly common to estimate an experiment's radiation and response characteristics with the help of computer simulations. General-purpose Monte-Carlo systems such as Geant4~\cite{2006ITNS...53..270A, Agostinelli2003250} are frequently the tool of choice for such simulations, as they allow for the modeling of complex geometries and a wide range of physical processes. The accuracy of these simulation-derived estimates is determined by the accuracy of the individual contributing processes, which in turn is validated by measurements.

The radioactive decay code implemented in the released version of Geant4 has been previously compared to experimental measurements in a number of works, e.g.~\cite{Hurtado5076086, 2004NIMPA.518..764H, Laborie2002618, Golovko2008266}. In these comparisons the simulated detector geometry is usually calibrated to the experiment in an iterative process so that the simulation better models the measurements. Whereas this method can produce simulation results consistent with experimental data within reasonable error margins determined by the experimental setup, it obfuscates how well a non-calibrated geometry can be simulated by Geant4. The ability to run absolute models is important if simulations are used to aid in the development of new detectors: in that context hardware, and thus measurements to compare against, do not exist.
 
A validation addressing this issue, like the one presented in this work, requires a self-consistent approach, i.e. only known experimental properties are used as simulation input. Accordingly, we have refrained from iteratively optimizing the detector geometry, with the sole exception of the simulated calibration source itself
 --- a topic discussed in Section~\ref{sec:sim_geom}.

The following sections provide an overview of the software for the simulation of radioactive decays in the Geant4 environment, discuss the strategy adopted for the validation process, and document the experimental measurements and the corresponding simulation. The validation results are divided into a comparison of the photo peaks in Section~\ref{sec:res_peak} and the complete spectra including the continuum in Section~\ref{sec:res_cont}.

\section{Radioactive Decays in Geant4}
A package for the simulation of radioactive decays~\cite{TruscottG4, 896281} has been available in the Geant4 simulation toolkit since version 2.0, in which it was named {\it radiative\_decay}. Since Geant4 version 6.0 it has been named {\it radioactive\_decay}, although it is conventionally known as Geant4 RDM (Radioactive Decay Module), and identified as ''original RDM'' in the following. Radioactive decays are simulated in this package by a process implemented in the G4RadioactiveDecay class, which samples secondary particles on a per-decay basis. The software implementation involves the optional use of event biasing techniques. The simulation of radioactive decays is data driven, using a reprocessed ENSDF library~\cite{ENSDF}. From this library the algorithm samples any direct decay emission (i.e. $\mathrm{\alpha}$ and $\mathrm{\beta}$ as well as neutrinos) resulting from nuclear transmutations. The sampling and production of deexcitation emission (i.e. $\mathrm{\gamma}$-rays and conversion electrons) is delegated to an independent Geant4 class: G4PhotonEvaporation. Similarly, the handling of fluorescence emission, resulting from the shell vacancies occurring from electron-capture decays, and if activated in the code, for $\mathrm{\beta}$-decays, is delegated to the independent G4AtomicDeexcitation class. Each of these three classes uses a separate set of data libraries, which may not necessarily be completely consistent with each other (e.g. with respect to energy).

An extended and improved package for Geant4-based simulation of radioactive decays, identified in the following as RDM-extended, has been developed. This package, which is described in detail in~\cite{RadDecay2012_1}, addresses some shortcomings identified in the original Geant4 RDM and extends its capabilities by providing the option of a novel statistical sampling approach along with a refactored version of the existing per-decay sampling.  The software design of the RDM-extended is compatible with the Geant4 kernel; therefore the original RDM and the RDM-extended can be used interchangeably in experimental simulation applications.

The refactored per-decay sampling of the RDM-extended package provides functionality equivalent to the original RDM. Additionally, as documented in~\cite{RadDecay2012_1}, the RDM-extended package provides more consistent production of conversion electrons and fluorescence emission, since all relevant data are obtained from a single library based on ENSDF. The novel statistical approach, described in detail in~\cite{RadDecay2012_1}, allows for the efficient simulation of a large number of decays. Here allowed or forbidden transitions during nuclear and atomic deexcitation are not taken into account for the individual decay; instead, the effect of these transitions on the radiation produced by many decays is considered. 

The public release of the RDM-extended package is foreseen following the publication of this paper; until it is publicly distributed, the software is available from the authors on request.

\section{Measurements}

\subsection{Experimental Setup}
\label{sec:exp_set}

\begin{figure}
\centering
\centerline{\includegraphics[width=3.4in]{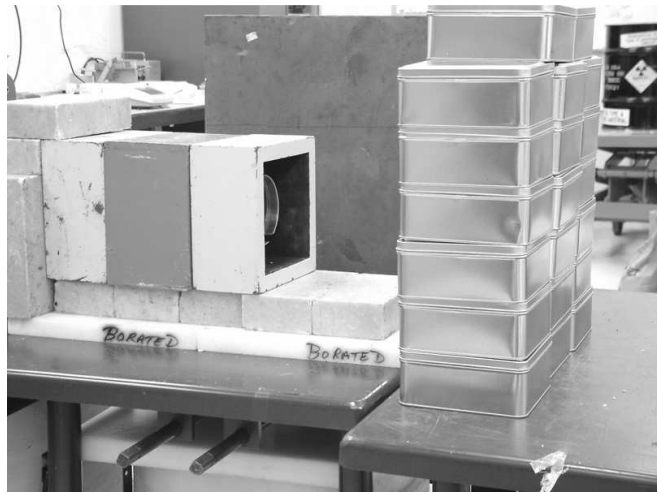}}
\caption{The experimental setup consisting of a HPGe detector, which is placed inside hollow lead shielding blocks shown in the center left part of the image. These blocks are the outer part of a collimator, which additionally consists of a tin and a copper tube, placed around the cylindrical detector head. Additional steel shielding can be seen on the tables. The source was suspended in front of the detector by a thin wire. Not visible is the detectors's dewar.}
\label{fig:pic_setup}
\end{figure}

\begin{figure}
\centering
\centerline{\includegraphics[width=3.in]{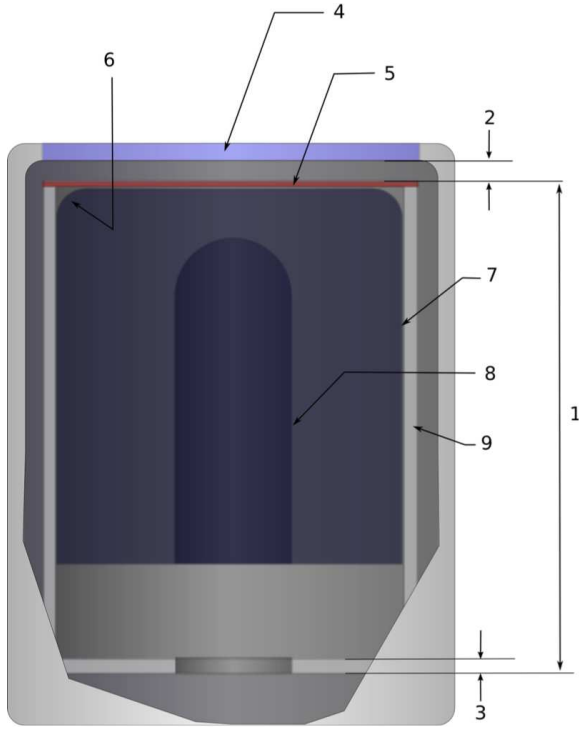}}
\caption{A sketch of the HPGE-Detector's head showing important measurements: the innerlying HPGe/crystal {\bf 1} mount cup length, {\bf 2} distance end cap - crystal, {\bf 3} mount cup base, {\bf 4} Beryllium entrance window, {\bf 5} Mylar insulator, {\bf 6} crystal end radius, {\bf 7} outer dead layer, {\bf 8} inner dead layer and bore, {\bf 9} mount cup thickness. Measurements were obtained from private communication with the manufacturer.}
\label{fig:detector_sketch}
\end{figure}

In order to minimize the experimental setup's uncertainties, we chose to use a simple setup which uses a High Purity Germanium detector (HPGe) for the detection of $\mathrm{\gamma}$-- and {X-rays} resulting from the nuclear and atomic deexcitation of the decaying calibration sources. A photograph of this setup is shown in Fig.~\ref{fig:pic_setup}. For the measurements we have used an ORTEC $70\,\mathrm{mm}$ diameter pop-top model detector. Its geometry, obtained from the manufacturer, is shown in Fig.~\ref{fig:detector_sketch}. The detector head was situated inside a lead shielding consisting of hollowed lead blocks with a bore of $16\times16\,\mathrm{cm}$ and $2\,\mathrm{cm}$ thick walls. This collimator additionally includes a copper ($\oslash=10.165\,\mathrm{cm}$, thickness $3.4\,\mathrm{mm}$) and a tin ($\oslash=10.3\,\mathrm{cm}$, thickness $1.35\,\mathrm{mm}$) tube, both of which are placed concentrically around the cylindrical detector head. The tubes are aligned with the dewar facing edge of the lead blocks and are $30.4\,\mathrm{cm}$ long. Additional shielding was provided by a steel block and steel-shot-filled boxes stacked on tables situated in front of the detector, which formed two perpendicular walls. The detector's dewar and the collimator's lead parts sat on top of borated wax blocks which were placed upon a wooden crate (dewar end) and the aforementioned steel table (collimator end).

Six different radioactive calibrations sources (Eckart \& Ziegler type D) containing the following isotopes were used: $\mathrm{^{22}Na}$, $\mathrm{^{54}Mn}$, $\mathrm{^{57}Co}$, $\mathrm{^{60}Co}$, $\mathrm{^{133}Ba}$, $\mathrm{^{137}Cs}$. The activities and decay characteristics of these sources are given in Table~\ref{tab:sources}, the measured intensities and energies of the prominent photo peaks are given in Table~\ref{tab:ensdf_comp}. The sources are enclosed in a plastic disc whose dimensions were obtained from the manufacturer ($\oslash=25.4\,\mathrm{mm}$, disc height: $6.25\,\mathrm{mm}$). The active deposit is located in a centrally placed hole with a diameter of $\oslash=5.0\,\mathrm{mm}$, a depth of $3.18\,\mathrm{mm}$ and covered by an expoxy plug. During the measurements these discs were suspended from a wire at a distance of $30\,\mathrm{cm}$ from the entrance window. Vertically the sources were centered on the detector's horizontal axis (i.e. the center axis of the crystal).

\begin{table}
\caption{Properties of the measured calibration sources. The half-life times were taken from the NuDat online database~\cite{sonzogni:574}.}
\label{tab:sources}
\centering{
\begin{tabular}{lrrrr}
\parbox{0.8cm}{Isotope} & \parbox{1.7cm}{\centering Reference \\ activity (date)} & \parbox{1.cm}{Half-life} & \parbox{1.5cm}{\centering Activity \\ meas. date} & \parbox{0.7cm}{Decay type} \\
\toprule
$\mathrm{^{22}Na}$ & $37.0\,\mathrm{kBq}$ (06/2006) & $2.603\,\mathrm{y}$ & $15.6\,\mathrm{kBq}$ & $\mathrm{\beta^{+}}$\\
$\mathrm{^{54}Mn}$ & $37.0\,\mathrm{kBq}$ (06/2006) & $312.120\,\mathrm{d}$ & $2.7\,\mathrm{kBq}$ & $\mathrm{EC}$\\
$\mathrm{^{57}Co}$ & $37.0\,\mathrm{kBq}$ (06/2006) & $271.740\,\mathrm{d}$ & $1.8\,\mathrm{kBq}$ & $\mathrm{EC}$\\
$\mathrm{^{60}Co}$ & $37.0\,\mathrm{kBq}$ (06/2006) & $1925.280\,\mathrm{d}$ & $24.1\,\mathrm{kBq}$ & $\mathrm{\beta^{-}}$\\
$\mathrm{^{133}Ba}$ & $37.0\,\mathrm{kBq}$ (05/2006) & $10.551\,\mathrm{y}$ & $29.7\,\mathrm{kBq}$ & $\mathrm{EC}$ \\
$\mathrm{^{137}Cs}$ & $37.0\,\mathrm{kBq}$ (06/2006) & $30.080\,\mathrm{y}$ & $34.3\,\mathrm{kBq}$ & $\mathrm{\beta^{-}}$\\
\bottomrule
\end{tabular}}
\end{table}

\begin{figure}
\centering
\centerline{\includegraphics[width=3.4in]{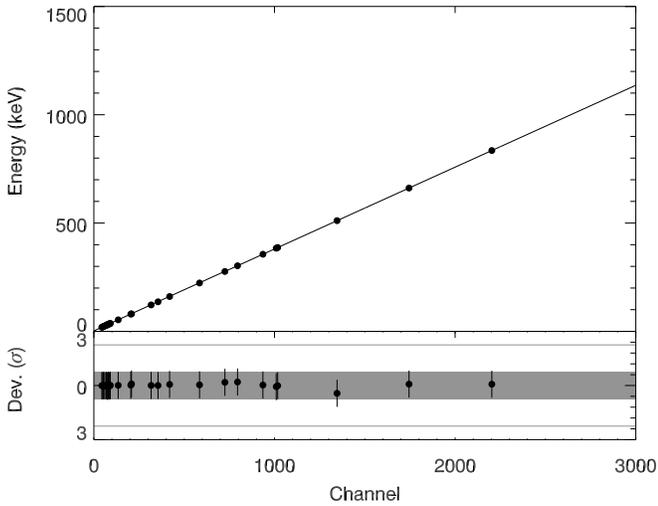}}
\caption{Energy calibration for the HPGe--detector: the upper part of the plot shows the photo peak locations determined with the {\it HYPERMET} program in comparison to reference data taken from NuDat (dots). The $1\mathrm{\sigma}$-uncertainties as determined by the fit are not distinguishable at this scale. Also shown is a polynomial fit to the data (solid line). The lower part shows the residuals in terms of $\mathrm{\sigma}$ uncertainties (solid area: $1\mathrm{\sigma}$, horizontal lines: $3\mathrm{\sigma}$).}
\label{fig:energy_fit}
\end{figure}

The individual isotope sources were measured for $10\,\mathrm{min}$ (live time) each. A background measurement of $60\,\mathrm{min}$ was performed prior to and after the complete set of isotope measurements. The mean of these two background spectra was subtracted from the source spectra before further analysis. Each spectrum was acquired with a binning of $8192\,\mathrm{channels}$. An energy calibration was performed by comparing the channel peak positions with reference data from the NuDat online database~\cite{sonzogni:574} (see Fig. \ref{fig:energy_fit}). 

\subsection{Experimental Data Analysis}
\label{sec:exp_analysis}

\begin{figure}
\centering
\centerline{\includegraphics[width=3.4in]{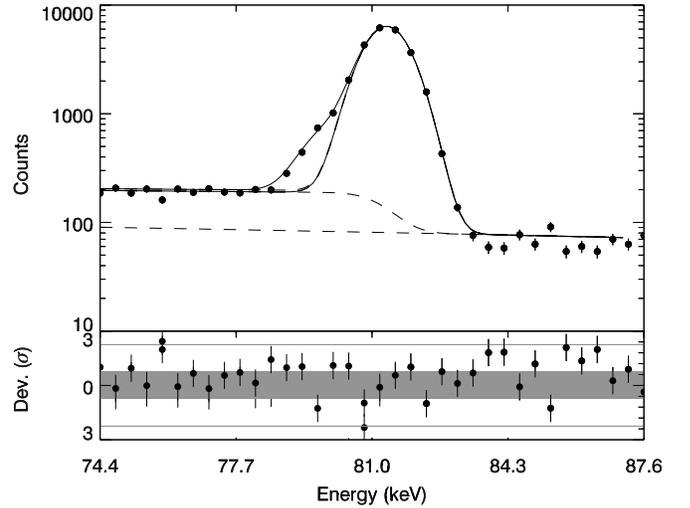}}
\caption{The upper part of the plot shows the measured $\mathrm{^{133}Ba}$ photo peaks at $79.61\,\mathrm{keV}$ and $80.99\,\mathrm{keV}$ (dots). Also shown are the background and peak models (described in Section~\ref{sec:exp_analysis}) fitted to the data using HYPERMET. The background consists of a constant linear part (lower dashed line) and a step-function modeling the Compton background component (upper dashed line) of the spectrum. The lower energy peak is shown dotted, whereas the solid line is the sum of the second peak and the other components. The low-energy exponential tail is not distinguishable for these peaks. The $1\mathrm{\sigma}$-uncertainties as determined by the fit are not always distinguishable at this scale. The lower part shows the residuals in terms of $\mathrm{\sigma}$ uncertainties (filled area: $1\mathrm{\sigma}$, horizontal lines: $3\mathrm{\sigma}$).}
\label{fig:peak_1}
\end{figure}

\begin{figure}
\centering
\centerline{\includegraphics[width=3.4in]{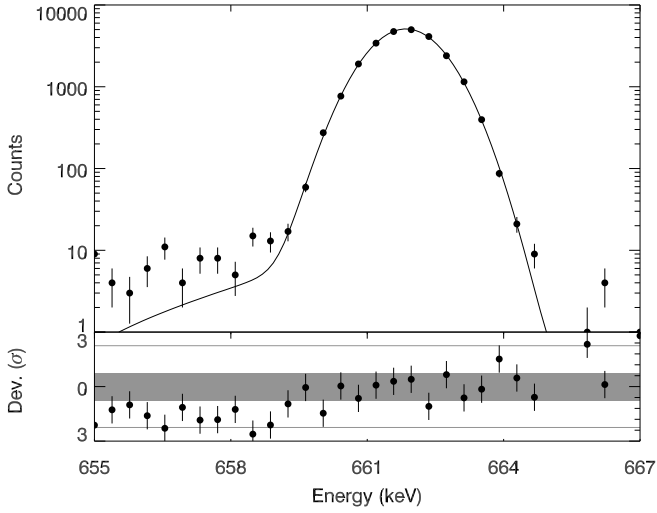}}
\caption{The upper part of the plot shows the measured $\mathrm{^{137}Cs}$ photo peak at $661.66\,\mathrm{keV}$ (dots). Also shown is the model fitted to the data using HYPERMET (solid line). Clearly visible is the exponential tail towards low energies, which is added to the Gaussian peak. The $1\mathrm{\sigma}$-uncertainties as determined by the fit are not always distinguishable at this scale. The lower part shows the residuals in terms of $\mathrm{\sigma}$ uncertainties (solid area: $1\mathrm{\sigma}$, horizontal lines: $3\mathrm{\sigma}$).}
\label{fig:peak_2}
\end{figure}

We have analyzed the experimental data using the {\it HYPERMET} program which models the spectral characteristics as given in~\cite{Phillips1976525}. This program has been chosen as it uses well-established models for both peak- and background components of $\mathrm{\gamma}$-ray decay spectra, which are based on semi-empirical functions as detailed in~\cite{Phillips1976525}. In this way we were assured that the peak and surrounding background characteristics were determined in a stable and proven fashion. To detail the fitting model needed for a precise analysis, a short summary of the individual model components for the peaks and background is presented in this section (in the order of decreasing importance of their impact on the spectrum). The expressions which model these components were used to model the detector response for the simulation. The prominent features described in the following are illustrated in Fig.~\ref{fig:peak_1} and~\ref{fig:peak_2}.

\begin{figure}
\centering
\centerline{\includegraphics[width=3.4in]{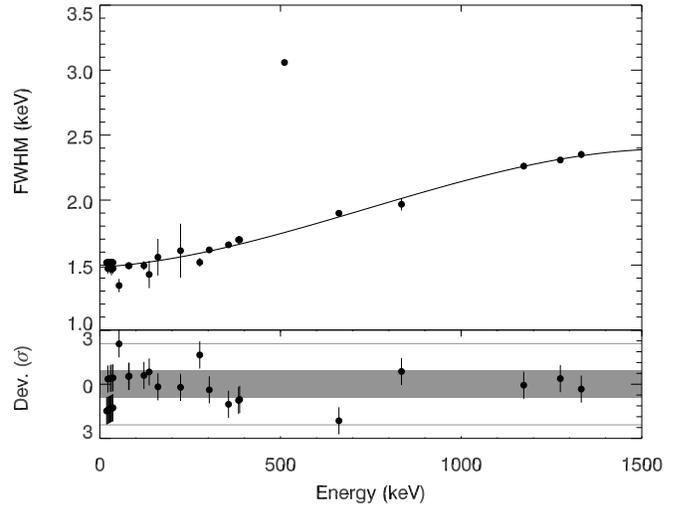}}
\caption{The upper part of the plot shows the measured photo peak FWHM for all isotopes (dots). The stray data point is the $511\,\mathrm{keV}$ pair-production peak, which has a much broader width than the photo peaks and was excluded from the fit. Also shown is a second degree polynomial model fitted to the data which was used to determine the detector resolution at arbitrary energies. The $1\mathrm{\sigma}$-uncertainties as determined by the fit are not always distinguishable at this scale. The lower part shows the residuals in terms of $\mathrm{\sigma}$ uncertainties (filled area: $1\mathrm{\sigma}$, horizontal lines: $3\mathrm{\sigma}$).}
\label{fig:fwhm_fit}
\end{figure}

\begin{figure}
\centering
\centerline{\includegraphics[width=3.4in]{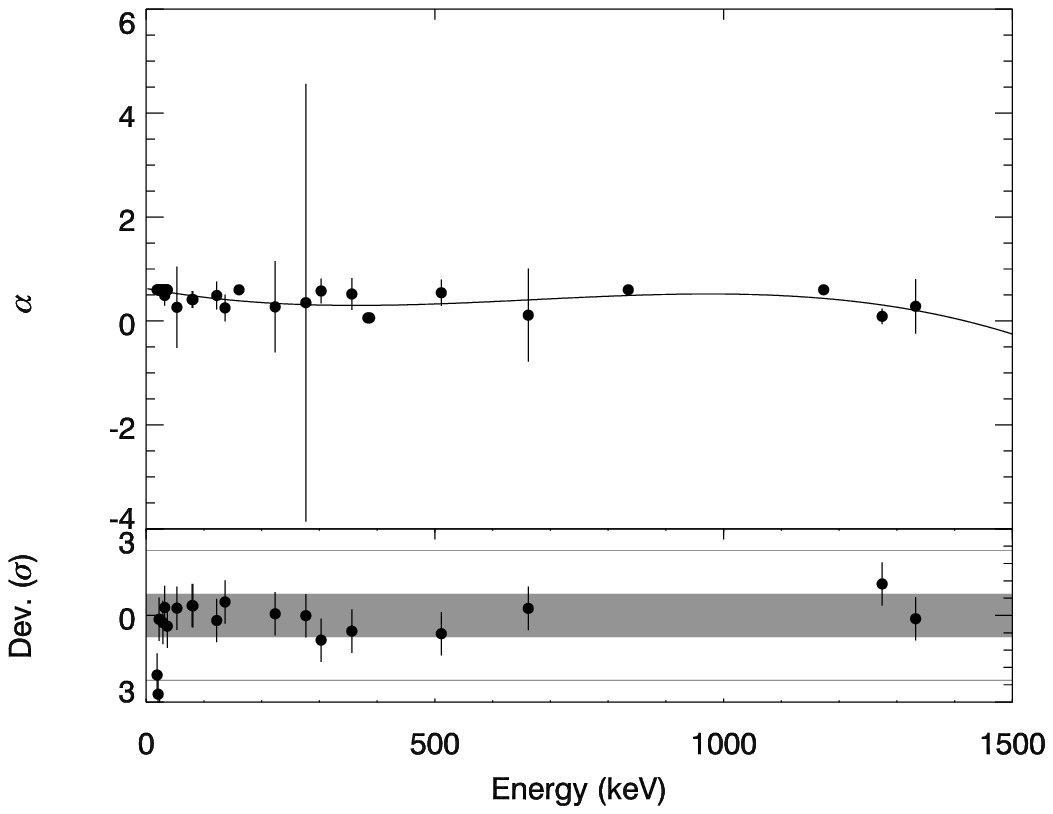}}
\caption{The upper part of the plot shows the $\alpha$ parameter defining the normalization of the exponential tail for the measured photo peaks(dots). Also shown is a third degree polynomial model fitted to the data, which was used to determine the parameter at arbitrary energies. The $1\mathrm{\sigma}$-uncertainties as determined by the fit are not always distinguishable at this scale. The lower part shows the residuals in terms of $\mathrm{\sigma}$ uncertainties (filled area: $1\mathrm{\sigma}$, horizontal lines: $3\mathrm{\sigma}$).}
\label{fig:alpha_fit}
\end{figure}

\begin{figure}
\centering
\centerline{\includegraphics[width=3.4in]{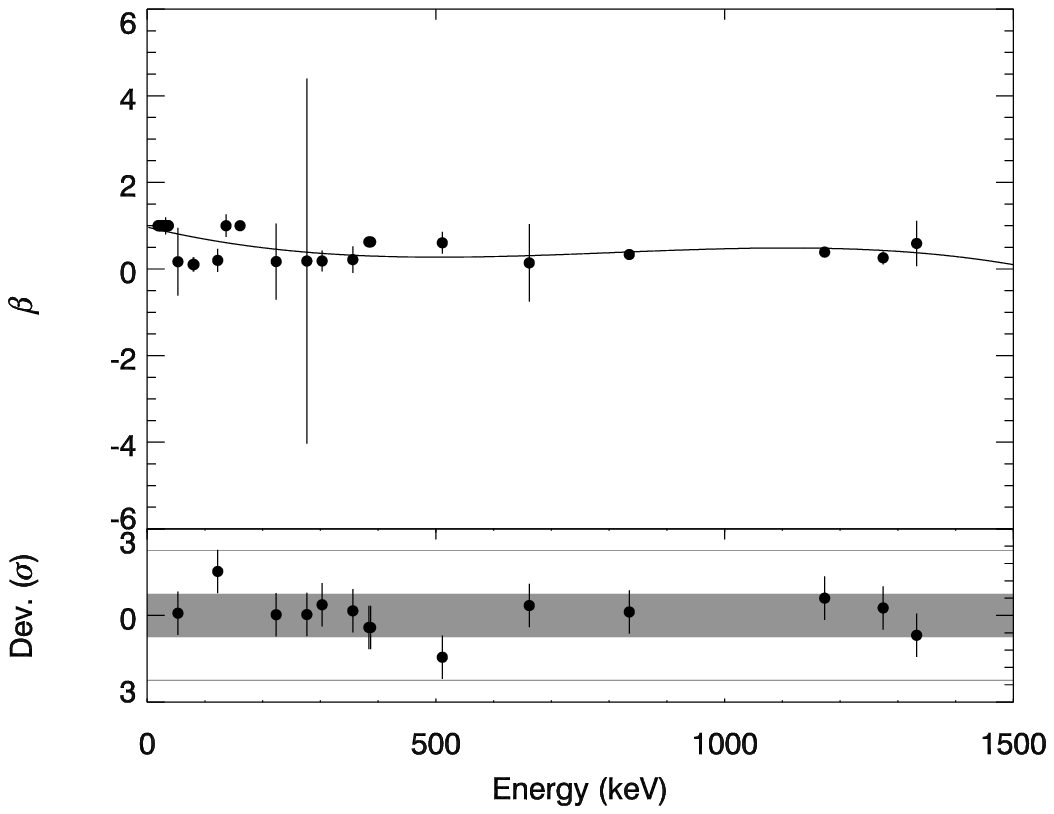}}
\caption{The upper part of the plot shows the $\beta$ parameter defining the slope of the exponential tail for the measured photo peaks(dots). Also shown is a third degree polynomial model fitted to the data, which was used to determine the parameter at arbitrary energies. The $1\mathrm{\sigma}$-uncertainties as determined by the fit are not always distinguishable at this scale. The lower part shows the residuals in terms of $\mathrm{\sigma}$ uncertainties (filled area: $1\mathrm{\sigma}$, horizontal lines: $3\mathrm{\sigma}$).}
\label{fig:beta_fit}
\end{figure}

\begin{enumerate}
\item Ideally, a $\mathrm{\gamma}$-ray registered in a detector would result in sharp spectral line at a single energy. Due to statistical effects and electronic noise the signal is instead broadened into a Gaussian distribution

\begin{equation}
A\,\mathrm{exp}{\left(-\frac{x-x_{c}}{\sigma}\right)}^{2}
\end{equation}

with amplitude $A$, channel $x$, center $x_{c}$ and the energy resolution defining width $\sigma$. For Germanium-detectors the width usually increases as the square of the $\mathrm{\gamma}$-ray energy, which is resembled in the polynomial fit shown in Fig.~\ref{fig:fwhm_fit}. Gaussian peak shapes are illustrated in Fig.~\ref{fig:peak_1} and~\ref{fig:peak_2}. 

\item Incomplete charge collection moves events from the photo peak towards lower energies. This results in an additional exponentially decaying component, $\mathrm{exp}(x-x_{c})/\beta)$, with slope $\beta$. Folded with a Gaussian noise of width $\gamma$ this yields 

\begin{equation}
\alpha\, \,\mathrm{exp}{\left(\frac{x-x_{c}}{\beta}\right)}\times \frac{1}{2} \mathrm{erfc}\left(\frac{x-x_{c}}{\gamma}+\frac{\gamma}{2\beta}\right)
\end{equation}

where $\alpha$ is a normalization factor to the amplitude $A$ and $x_{c}$ is the position of the photo peak's center; $\mathrm{erfc}(u)$ is the complementary error function. As the parameters $\alpha$ and $\beta$ may vary with energy, the polynomial models shown in Fig.~\ref{fig:alpha_fit} and~\ref{fig:beta_fit} are used to determine their magnitude at arbitrary energies. Fig.~\ref{fig:peak_2} shows as an example the strong exponential component of the $\mathrm{^{137}Cs}$ photo peak at $661.66\,\mathrm{keV}$, which is underestimated by the model below $658\,\mathrm{keV}$. 

\item The most intensive photo peaks may exhibit an additional, longer exponential tail at energies below the peak, which can be attributed to surface effects and will also influence lower lying peaks. It has the same form as the exponential tail (2), but with different parameters: 

\begin{equation}
\tau\,A\,\mathrm{exp}{\left(\frac{x-x_{c}}{\nu}\right)}\times \frac{1}{2} \mathrm{erfc}\left(\frac{x-x_{c}}{\mu}+\frac{\mu}{2\nu}\right)
\end{equation}

The slope $\nu$ is usually one to two orders of magnitude larger than $\beta$, whereas the relative amplitude $\tau$ is one to three orders smaller~\cite{Phillips1976525}. The width of the Gaussian noise is determined by $\mu$. 

\item Within the detector $\mathrm{\gamma}$-rays can Compton-scatter and also exit the sensitive volume of the crystal. These events result in a ledge in the background above or below the photo peaks. Such a step of amplitude $S$ folded with Gaussian noise of width $\delta$ can be modeled using

\begin{equation}
S\times \frac{1}{2}\mathrm{erfc}\left(\frac{x-x_{c}}{\delta}\right).
\end{equation}

\item A nearly linear background exists within a few peak widths on either side of a photo peak. It can be mathematically represented by a polynomial function

\begin{equation}
B(x) = a+b\,\left(x-x_{c}\right)+c\,{\left(x-x_{c}\right)}^2.
\end{equation}

The total background underlying a specific peak is given by the sum of components 3 to 5. The background from peaks at higher energies may contribute to the background of those peaks at lower energies. On top of this sum of backgrounds the peak itself (component 1) and the corresponding exponential tail (component 2) have to be added.

Table~\ref{tab:parameters_1} summarizes the parameters discussed above as determined by fitting the experimental data with HYPERMET, whereby $\gamma$, $\delta$ and $\mu$ are set to the best-fit $\sigma$-value by the program. The $511\,\mathrm{keV}$ line of $\mathrm{^{22}Na}$ is broadened due to kinematic effects when the positrons annihilate in the source. It was excluded from the data used for the fit which determines the line broadening for the other photo-peaks. Accordingly, the peak width for the simulated $\mathrm{^{22}Na}$ was determined directly from the experimental peak width.

\end{enumerate}

\begin{table*}
\caption{Peak and background parameters obtained by fitting the experimental data with the HYPERMET program}
\label{tab:parameters_1}
\centerline{
\begin{tabular}{lrrrrrrr}%rrrrrrrr}
& \multicolumn{1}{c}{E [keV]} & \multicolumn{1}{c}{Channel} & \multicolumn{1}{c}{$\mathrm{A}$} & \multicolumn{1}{c}{$\mathrm{\sigma}$} & \multicolumn{1}{c}{Area} & \multicolumn{1}{c}{$\mathrm{\alpha}$} & \multicolumn{1}{c}{$\mathrm{\beta}$}\\
\toprule
$\mathrm{^{22}Na}$ & $1274.55 \pm 0.10 $ & $ 3366.57 \pm 0.10 $ & $ 2041 \pm \Z20 $ & $ 5.44 \pm 0.04 $ & $ 11812 \pm \Z105 $ & $ 0.09 \pm 0.15 $ & $ 0.26 \pm 0.49 $\medskip\\
$\mathrm{^{54}Mn}$ & $834.87 \pm 0.06 $ & $ 2203.08 \pm 0.09 $ & $ 411 \pm \Z26 $ & $ 4.63 \pm 0.11 $ & $ 2081 \pm \Z\Z54 $ & $ 0.60 \pm 0.01 $ & $ 0.34 \pm 0.36 $\medskip\\
$\mathrm{^{57}Co}$ & $122.06 \pm 0.03 $ & $ 317.05 \pm 0.04 $ & $ 763 \pm \Z23 $ & $ 3.52 \pm 0.08 $ & $ 2863 \pm \Z\Z80 $ & $ 0.49 \pm 0.27 $ & $ 0.20 \pm 0.22 $ \\
    & $136.48 \pm 0.05 $ & $ 355.20 \pm 0.13 $ & $ 85 \pm \Z\Z8 $ & $ 3.63 \pm 0.25 $ & $ 338 \pm \Z\Z22 $ & $ 0.25 \pm 0.26 $ & $ 1.00 \pm 0.01 $\medskip\\
$\mathrm{^{60}Co}$ & $1174.23 \pm 0.07 $ & $ 3098.50 \pm 0.04 $ & $ 1996 \pm \Z45 $ & $ 5.33 \pm 0.06 $ & $ 11603 \pm \Z112 $ & $ 0.60 \pm 0.01 $ & $ 0.39 \pm 0.11 $ \\
    & $1332.50 \pm 0.09 $ & $ 3520.01 \pm 0.05 $ & $ 1748 \pm \Z64 $ & $ 5.54 \pm 0.07 $ & $ 10770 \pm \Z117 $ & $ 0.28 \pm 0.52 $ & $ 0.59 \pm 0.23 $\medskip\\
$\mathrm{^{133}Ba}$ & $ 27.98 \pm 0.11 $ & $ 68.16 \pm 0.29 $ & $ 301 \pm \Z33 $ & $ 3.58 \pm 0.04 $ & $ 1446 \pm \Z154 $ & $ 0.60 \pm 0.00 $ & $ 1.00 \pm 0.00 $ \\
    & $ 30.77 \pm 0.02 $ & $ 75.54 \pm 0.02 $ & $ 11931 \pm 204 $ & $ 3.58 \pm 0.04 $ & $ 57391 \pm 1061 $ & $ 0.60 \pm 0.00 $ & $ 1.00 \pm 0.00 $ \\
    & $ 32.66 \pm 0.28 $ & $ 80.54 \pm 0.74 $ & $ 121 \pm \Z34 $ & $ 3.58 \pm 0.04 $ & $ 582 \pm \Z161 $ & $ 0.60 \pm 0.00 $ & $ 1.00 \pm 0.00 $ \\
    & $ 35.00 \pm 0.03 $ & $ 86.72 \pm 0.07 $ & $ 2793 \pm 113 $ & $ 3.58 \pm 0.04 $ & $ 13437 \pm \Z471 $ & $ 0.60 \pm 0.00 $ & $ 1.00 \pm 0.00 $ \\
    & $ 36.10 \pm 0.15 $ & $ 89.65 \pm 0.40 $ & $ 265 \pm \Z78 $ & $ 3.58 \pm 0.04 $ & $ 1272 \pm \Z373 $ & $ 0.60 \pm 0.00 $ & $ 1.00 \pm 0.00 $ \\
    & $ 53.16 \pm 0.03 $ & $ 134.77 \pm 0.05 $ & $ 410 \pm \Z18 $ & $ 3.16 \pm 0.12 $ & $ 1380 \pm \Z\Z52 $ & $ 0.26 \pm 0.78 $ & $ 0.17 \pm 6.17 $ \\
    & $ 79.53 \pm 0.05 $ & $ 204.55 \pm 0.13 $ & $ 455 \pm \Z38 $ & $ 3.52 \pm 0.03 $ & $ 1706 \pm \Z131 $ & $ 0.41 \pm 0.16 $ & $ 0.10 \pm 0.02 $ \\
    & $ 81.01 \pm 0.02 $ & $ 208.45 \pm 0.02 $ & $ 6251 \pm \Z68 $ & $ 3.52 \pm 0.03 $ & $ 23428 \pm \Z232 $ & $ 0.41 \pm 0.16 $ & $ 0.10 \pm 0.02 $ \\
    & $ 160.78 \pm 0.06 $ & $ 419.49 \pm 0.14 $ & $ 94 \pm \Z\Z8 $ & $ 3.68 \pm 0.33 $ & $ 466 \pm \Z\Z43 $ & $ 0.60 \pm 0.00 $ & $ 1.00 \pm 0.00 $ \\
    & $ 223.32 \pm 0.10 $ & $ 584.95 \pm 0.28 $ & $ 64 \pm \Z10 $ & $ 3.79 \pm 0.49 $ & $ 257 \pm \Z\Z43 $ & $ 0.27 \pm 0.88 $ & $ 0.17 \pm 8.02 $ \\
    & $ 276.41 \pm 0.02 $ & $ 725.41 \pm 0.04 $ & $ 876 \pm \Z26 $ & $ 3.58 \pm 0.08 $ & $ 3342 \pm \Z\Z82 $ & $ 0.35 \pm 4.21 $ & $ 0.18 \pm 4.83 $ \\
    & $ 302.85 \pm 0.01 $ & $ 795.36 \pm 0.03 $ & $ 1951 \pm \Z45 $ & $ 3.81 \pm 0.06 $ & $ 7909 \pm \Z144 $ & $ 0.58 \pm 0.24 $ & $ 0.19 \pm 0.36 $ \\
    & $ 356.01 \pm 0.01 $ & $ 936.01 \pm 0.02 $ & $ 5965 \pm \Z69 $ & $ 3.90 \pm 0.03 $ & $ 24774 \pm \Z234 $ & $ 0.52 \pm 0.30 $ & $ 0.22 \pm 0.48 $ \\
    & $ 383.84 \pm 0.02 $ & $ 1009.62 \pm 0.04 $ & $ 793 \pm \Z20 $ & $ 3.99 \pm 0.07 $ & $ 3407 \pm \Z\Z68 $ & $ 0.06 \pm 0.00 $ & $ 0.63 \pm 0.58 $ \\
    & $ 386.89 \pm 0.11 $ & $ 1017.68 \pm 0.28 $ & $ 19 \pm \Z\Z2 $ & $ 3.99 \pm 0.07 $ & $ 83 \pm \Z\Z11 $ & $ 0.06 \pm 0.00 $ & $ 0.63 \pm 0.58 $\medskip\\
$\mathrm{^{137}Cs}$ & $ 29.01 \pm 0.20 $ & $ 70.89 \pm 0.51 $ & $ 30 \pm \Z\Z7 $ & $ 3.47 \pm 0.09 $ & $ 139 \pm \Z\Z29 $ & $ 0.60 \pm 0.08 $ & $ 1.00 \pm 0.00 $ \\
    & $ 32.01 \pm 0.04 $ & $ 78.83 \pm 0.07 $ & $ 878 \pm \Z43 $ & $ 3.46 \pm 0.11 $ & $ 3929 \pm \Z\Z86 $ & $ 0.49 \pm 0.20 $ & $ 1.00 \pm 0.02 $ \\
    & $ 36.50 \pm 0.05 $ & $ 90.70 \pm 0.11 $ & $ 191 \pm \Z11 $ & $ 3.47 \pm 0.09 $ & $ 890 \pm \Z\Z49 $ & $ 0.60 \pm 0.08 $ & $ 1.00 \pm 0.00 $ \\
    & $ 661.65 \pm 0.03 $ & $ 1744.69 \pm 0.02 $ & $ 5088 \pm \Z46 $ & $ 4.47 \pm 0.03 $ & $ 24222 \pm \Z198 $ & $ 0.11 \pm 0.90 $ & $ 0.14 \pm 0.37 $ \\
\bottomrule
\end{tabular}}
\vspace{1cm}‚
\centerline{
\begin{tabular}{lrrrrrrrr}%rrrrrrrr}
& \multicolumn{1}{c}{E [keV]} & \multicolumn{1}{c}{$\mathrm{a}$} & \multicolumn{1}{c}{$\mathrm{b}$} & \multicolumn{1}{c}{$\mathrm{c}$} & \multicolumn{1}{c}{$\mathrm{\Sigma}$} & \multicolumn{1}{c}{$\mathrm{\tau}$} & \multicolumn{1}{c}{$\mathrm{\nu}$} & \multicolumn{1}{c}{$\mathrm{\chi^{2}/d.o.f}$}\\
\toprule
$\mathrm{^{22}Na}$ & $ 1274.55 \pm 0.10 $ & $ 0.01 \pm 0.15 $ & $ 0.01 \pm 0.01 $ & $ 0.00 \pm 0.00 $ & $ 0.00 \pm 0.00 $ & $ 0.01 \pm 0.01 $ & $ 2.78 \pm 1.48 $ & $ 0.9 $\medskip\\
$\mathrm{^{54}Mn}$ & $ 834.87 \pm 0.06 $ & $ 0.60 \pm 0.45 $ & $ -0.01 \pm 0.02 $ & $ 0.00 \pm 0.00 $ & $ 0.00 \pm 0.00 $ & $ 0.00 \pm 0.00 $ & $ 0.00 \pm 0.00 $ & $ 1.3 $\medskip\\
$\mathrm{^{57}Co}$ & $ 122.06 \pm 0.03 $ & $ 0.00 \pm 0.01 $ & $ 0.04 \pm 0.04 $ & $ 0.00 \pm 0.00 $ & $ 0.01 \pm 0.00 $ & $ 0.00 \pm 0.00 $ & $ 0.00 \pm 0.00 $ & $ 2.2 $\\
    & $ 136.48 \pm 0.05 $ & $ 0.94 \pm 0.73 $ & $ 0.01 \pm 0.04 $ & $ 0.00 \pm 0.00 $ & $ 0.00 \pm 0.00 $ & $ 0.00 \pm 0.00 $ & $ 0.00 \pm 0.00 $ & $ 1.5 $\medskip\\
$\mathrm{^{60}Co}$ & $ 1174.23 \pm 0.07 $ & $ 2.81 \pm 9.98 $ & $ 0.27 \pm 0.34 $ & $ 0.00 \pm 0.00 $ & $ 0.01 \pm 0.00 $ & $ 0.00 \pm 0.00 $ & $ 11.73 \pm 20.51 $ & $ 1.1 $\\
    & $ 1332.50 \pm 0.09 $ & $ 3.17 \pm 0.80 $ & $ -0.06 \pm 0.03 $ & $ 0.00 \pm 0.00 $ & $ 0.01 \pm 0.00 $ & $ 0.00 \pm 0.00 $ & $ 0.00 \pm 0.00 $ & $ 1.3 $\medskip\\
$\mathrm{^{133}Ba}$ & $ 27.98 \pm 0.11 $ & $ 335.19 \pm 7.82 $ & $ -4.28 \pm 0.22 $ & $ 0.00 \pm 0.00 $ & $ -0.02 \pm 0.00 $ & $ 0.00 \pm 0.00 $ & $ 1.00 \pm 0.00 $ & $ 8.6 $ \\
    & $ 30.77 \pm 0.02 $ & $ 335.19 \pm 7.82 $ & $ -4.28 \pm 0.22 $ & $ 0.00 \pm 0.00 $ & $ -0.02 \pm 0.00 $ & $ 0.00 \pm 0.00 $ & $ 1.00 \pm 0.00 $ & $ 8.6 $ \\
    & $ 32.66 \pm 0.28 $ & $ 335.19 \pm 7.82 $ & $ -4.28 \pm 0.22 $ & $ 0.00 \pm 0.00 $ & $ -0.02 \pm 0.00 $ & $ 0.00 \pm 0.00 $ & $ 1.00 \pm 0.00 $ & $ 8.6 $ \\
    & $ 35.00 \pm 0.03 $ & $ 335.19 \pm 7.82 $ & $ -4.28 \pm 0.22 $ & $ 0.00 \pm 0.00 $ & $ -0.02 \pm 0.00 $ & $ 0.00 \pm 0.00 $ & $ 1.00 \pm 0.00 $ & $ 8.6 $ \\
    & $ 36.10 \pm 0.15 $ & $ 335.19 \pm 7.82 $ & $ -4.28 \pm 0.22 $ & $ 0.00 \pm 0.00 $ & $ -0.02 \pm 0.00 $ & $ 0.00 \pm 0.00 $ & $ 1.00 \pm 0.00 $ & $ 8.6 $ \\
    & $ 53.16 \pm 0.03 $ & $ 70.32 \pm 3.57 $ & $ 0.48 \pm 0.21 $ & $ 0.00 \pm 0.00 $ & $ 0.00 \pm 0.00 $ & $ 0.00 \pm 0.00 $ & $ 0.00 \pm 0.00 $ & $ 1.4 $ \\
    & $ 79.53 \pm 0.05 $ & $ 80.60 \pm 23.96 $ & $ -0.53 \pm 0.77 $ & $ 0.00 \pm 0.00 $ & $ 0.02 \pm 0.00 $ & $ 0.00 \pm 0.00 $ & $ 1.00 \pm 0.00 $ & $ 1.7 $ \\
    & $ 81.01 \pm 0.02 $ & $ 80.60 \pm 23.96 $ & $ -0.53 \pm 0.77 $ & $ 0.00 \pm 0.00 $ & $ 0.02 \pm 0.00 $ & $ 0.00 \pm 0.00 $ & $ 1.00 \pm 0.00 $ & $ 1.7 $ \\
    & $ 160.78 \pm 0.06 $ & $ 127.26 \pm 4.08 $ & $ -0.68 \pm 0.31 $ & $ 0.00 \pm 0.00 $ & $ 0.00 \pm 0.00 $ & $ 0.00 \pm 0.00 $ & $ 0.00 \pm 0.00 $ & $ 0.7 $ \\
    & $ 223.32 \pm 0.10 $ & $ 70.11 \pm 4.43 $ & $ -0.08 \pm 0.32 $ & $ 0.00 \pm 0.00 $ & $ 0.00 \pm 0.00 $ & $ 0.00 \pm 0.00 $ & $ 0.00 \pm 0.00 $ & $ 1.5 $ \\
    & $ 276.41 \pm 0.02 $ & $ 54.62 \pm 3.25 $ & $ -0.11 \pm 0.20 $ & $ 0.00 \pm 0.00 $ & $ 0.00 \pm 0.00 $ & $ 0.00 \pm 0.00 $ & $ 0.00 \pm 0.00 $ & $ 1.8 $ \\
    & $ 302.85 \pm 0.01 $ & $ 43.40 \pm 5.86 $ & $ -0.05 \pm 0.21 $ & $ 0.00 \pm 0.00 $ & $ 0.00 \pm 0.00 $ & $ 0.01 \pm 0.01 $ & $ 2.50 \pm 0.09 $ & $ 2.3 $ \\
    & $ 356.01 \pm 0.01 $ & $ 9.89 \pm 5.25 $ & $ -0.13 \pm 0.19 $ & $ 0.00 \pm 0.00 $ & $ 0.00 \pm 0.00 $ & $ 0.00 \pm 0.00 $ & $ 5.23 \pm 4.59 $ & $ 2.2 $ \\
    & $ 383.84 \pm 0.02 $ & $ 0.00 \pm 0.01 $ & $ 0.02 \pm 0.02 $ & $ 0.00 \pm 0.00 $ & $ 0.01 \pm 0.00 $ & $ 0.00 \pm 0.00 $ & $ 1.00 \pm 0.00 $ & $ 1.3 $ \\
    & $ 386.89 \pm 0.11 $ & $ 0.00 \pm 0.01 $ & $ 0.02 \pm 0.02 $ & $ 0.00 \pm 0.00 $ & $ 0.01 \pm 0.00 $ & $ 0.00 \pm 0.00 $ & $ 1.00 \pm 0.00 $ & $ 1.3 $\medskip\\
$\mathrm{^{137}Cs}$ & $ 29.01 \pm 0.20 $ & $ 33.56 \pm 2.35 $ & $ 0.09 \pm 0.06 $ & $ 0.00 \pm 0.00 $ & $ 0.00 \pm 0.00 $ & $ 0.00 \pm 0.00 $ & $ 0.00 \pm 0.00 $ & $ 1.7 $ \\
    & $ 32.01 \pm 0.04 $ & $ 42.56 \pm 3.77 $ & $ -0.08 \pm 0.14 $ & $ 0.00 \pm 0.00 $ & $ 0.00 \pm 0.00 $ & $ 0.00 \pm 0.00 $ & $ 0.00 \pm 0.00 $ & $ 1.7 $ \\
    & $ 36.50 \pm 0.05 $ & $ 33.56 \pm 2.35 $ & $ 0.09 \pm 0.06 $ & $ 0.00 \pm 0.00 $ & $ 0.00 \pm 0.00 $ & $ 0.00 \pm 0.00 $ & $ 0.00 \pm 0.00 $ & $ 1.7 $ \\
    & $ 661.65 \pm 0.03 $ & $ 0.00 \pm 0.01 $ & $ 0.03 \pm 0.02 $ & $ 0.00 \pm 0.00 $ & $ 0.00 \pm 0.00 $ & $ 0.00 \pm 0.00 $ & $ 7.14 \pm 5.26 $ & $ 1.6 $ \\
\bottomrule
\end{tabular}}
\end{table*}

\begin{table*}[!h]
\caption{Verification results for per-decay sampling (PD) and the RDM-extended statistical sampling (EXT) with respect to ENSDF evaluated data. Uncertainties of $0.00$ denote that uncertainties are present, but below the two significant digits given. The intensity deviation is given by $(I_{\mathrm{ref}}-I_{\mathrm{sim}})/I_{\mathrm{ref}}$.}
\label{tab:ensdf_comp}
\centerline{
\scriptsize 
\begin{tabular}{lrrrrrrr}
Isotope & Energy(keV) & Intensity (\%) & PD energy dev. (keV) & PD intensity dev. & EXT energy dev. (keV) & EXT intensity dev. \\
\toprule
$\mathrm{^{22}Na}$  & $1274.54 \pm 0.01$ & $99.94 \pm 1.40 \times 10^{-02}$ & $(1.37 \pm 2.00) \times 10^{-01}$ & $-0.00 \pm 0.00$ & $(1.37 \pm 2.00) \times 10^{-01}$ & $0.00 \pm 0.00$ \\
\midrule
$\mathrm{^{54}Mn}$ & $834.85 \pm 0.00$ & $99.98 \pm 1.00 \times 10^{-02}$ & $(0.48 \pm 2.00) \times 10^{-01}$ & $0.00 \pm 0.00$ & $0.23 \pm 2.00 \times 10^{-01}$ & $0.00 \pm 0.00$ \\
\midrule
$\mathrm{^{57}Co}$ & $14.41 \pm 0.00$ & $9.16 \pm 1.50 \times 10^{-01}$ & $(1.29 \pm 5.00) \times 10^{-02}$ & $0.00 \pm 0.00$ & $(1.29 \pm 5.00) \times 10^{-02}$ & $0.00 \pm 0.00$ \\
   & $122.06 \pm 0.00$ & $85.60 \pm 1.70 \times 10^{+00}$ & $(0.61 \pm 2.00) \times 10^{-01}$ & $-0.00 \pm 0.00$ & $(0.61 \pm 2.00) \times 10^{-01}$ & $-0.00 \pm 0.00$ \\
   & $136.47 \pm 0.00$ & $10.68 \pm 8.00 \times 10^{-02}$ & $(0.74 \pm 2.00) \times 10^{-01}$ & $0.03 \pm 0.00$ & $(0.74 \pm 2.00) \times 10^{-01}$ & $0.00 \pm 0.00$ \\
\midrule
$\mathrm{^{60}Co}$ & $1173.23 \pm 0.00$ & $99.85 \pm 3.00 \times 10^{-02}$ & $(0.28 \pm 2.00) \times 10^{-01}$ & $-0.00 \pm 0.00$ & $(0.13 \pm 2.00) \times 10^{-01}$ & $-0.00 \pm 0.00$ \\
   & $1332.49 \pm 0.00$ & $99.98 \pm 6.00 \times 10^{-04}$ & $(0.92 \pm 2.00) \times 10^{-01}$ & $-0.00 \pm 0.00$ & $(0.18 \pm 2.00) \times 10^{-01}$ & $-0.00 \pm 0.00$ \\
\midrule
$\mathrm{^{137}Cs}$ & $283.50 \pm 1.00$ & $(0.00 \pm 0.00) \times 10^{+00}$ & $(0.10 \pm 1.02) \times 10^{+00}$ & $0.14 \pm 0.00$ & $(0.10 \pm 1.02) \times 10^{+00}$ & $0.01 \pm 0.00$ \\
   & $661.66 \pm 0.00$ & $85.10 \pm 2.00 \times 10^{+00}$ & $(0.57 \pm 2.00) \times 10^{-01}$ & $0.00 \pm 0.00$ & $(0.57 \pm 2.00) \times 10^{-01}$ & $-0.00 \pm 0.00$ \\
\midrule
$\mathrm{^{133}Ba}$ & $30.63 \pm 0.00$ & $33.70 \pm 1.00 \times 10^{+00}$ & $(-1.75 \pm 2.00) \times 10^{-01}$ & $-0.52 \pm 0.00$ & $(-1.75 \pm 2.00) \times 10^{-01}$ & $-0.01 \pm 0.00$ \\
   & $30.97 \pm 0.00$ & $62.20 \pm 1.80 \times 10^{+00}$ & $(1.73 \pm 2.00) \times 10^{-01}$ & $0.17 \pm 0.00$ & $(1.73 \pm 2.00) \times 10^{-01}$ & $-0.02 \pm 0.00$ \\
   & $35.00 \pm \,\;-\;\;$ & $22.60 \pm 7.00 \times 10^{-01}$ & $(0.20 \pm \,\;-\;\;) \times 10^{+00}$ & $0.59 \pm 0.00$ & $(0.00 \pm \,\;-\;\;) \times 10^{+00}$ & $-0.05 \pm 0.00$ \\
   & $53.16 \pm 0.00$ & $2.20 \pm 2.20 \times 10^{-02}$ & $(1.62 \pm 2.00) \times 10^{-01}$ & $0.04 \pm 0.00$ & $(1.62 \pm 2.00) \times 10^{-01}$ & $-0.01 \pm 0.00$ \\
   & $79.61 \pm 0.00$ & $2.62 \pm 6.00 \times 10^{-02}$ & $(0.14 \pm 2.00) \times 10^{-01}$ & $0.01 \pm 0.00$ & $(0.14 \pm 2.00) \times 10^{-01}$ & $-0.00 \pm 0.00$ \\
   & $81.00 \pm 0.00$ & $34.10 \pm 3.00 \times 10^{-01}$ & $(1.97 \pm 2.00) \times 10^{-01}$ & $0.02 \pm 0.00$ & $(1.97 \pm 2.00) \times 10^{-01}$ & $0.00 \pm 0.00$ \\
   & $160.61 \pm 0.00$ & $(645.00 \pm 8.00) \times 10^{-03}$ & $(0.11 \pm 2.00) \times 10^{-01}$ & $0.03 \pm 0.00$ & $(0.11 \pm 2.00) \times 10^{-01}$ & $0.01 \pm 0.00$ \\
   & $223.24 \pm 0.00$ & $(45.00 \pm 4.00) \times 10^{-02}$ & $(0.37 \pm 2.00) \times 10^{-01}$ & $0.03 \pm 0.00$ & $(0.37 \pm 2.00) \times 10^{-01}$ & $0.02 \pm 0.00$ \\
   & $276.40 \pm 0.00$ & $7.16 \pm 2.20 \times 10^{-02}$ & $(2.00 \pm 2.00) \times 10^{-01}$ & $0.04 \pm 0.00$ & $(2.00 \pm 2.00) \times 10^{-01}$ & $-0.00 \pm 0.00$ \\
   & $302.85 \pm 0.01$ & $18.33 \pm 6.00 \times 10^{-02}$ & $(0.51 \pm 2.00) \times 10^{-01}$ & $0.02 \pm 0.00$ & $(0.51 \pm 2.00) \times 10^{-01}$ & $-0.00 \pm 0.00$ \\
   & $356.01 \pm 0.00$ & $62.05 \pm 1.90 \times 10^{-01}$ & $(0.13 \pm 2.00) \times 10^{-01}$ & $0.04 \pm 0.00$ & $(0.13 \pm 2.00) \times 10^{-01}$ & $-0.00 \pm 0.00$ \\
   & $383.85 \pm 0.01$ & $8.94 \pm 3.00 \times 10^{-03}$ & $(0.48 \pm 2.00) \times 10^{-01}$ & $0.02 \pm 0.00$ & $(0.48 \pm 2.00) \times 10^{-01}$ & $0.00 \pm 0.00$ \\
\bottomrule
\end{tabular}
}
\end{table*}

\begin{figure*}[!h]
\centerline{\subfloat[$\mathrm{^{22}Na}$]{\includegraphics[width=3.4in]{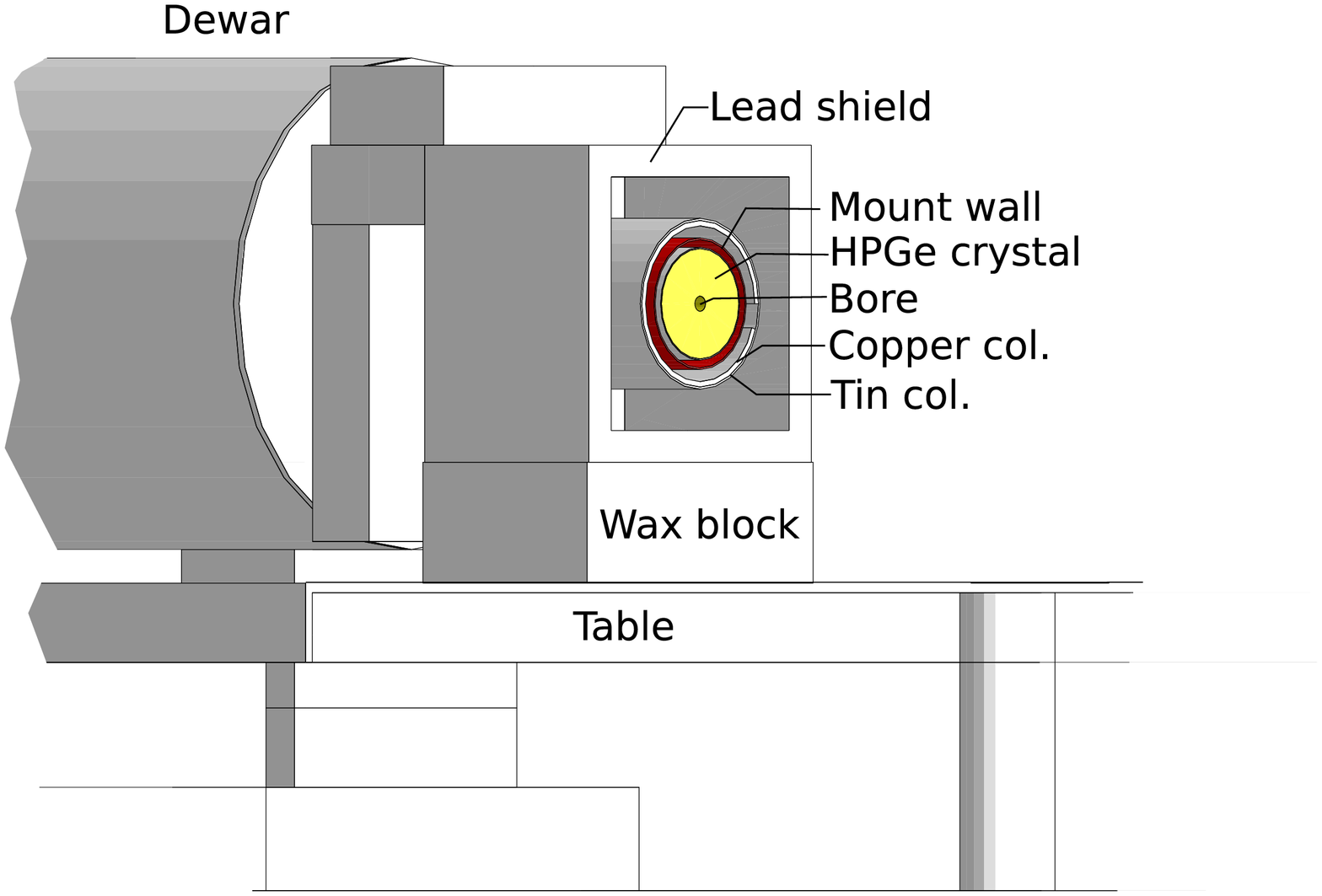}%
\label{fig:sim_setup1}}
\hfil
\subfloat[$\mathrm{^{54}Mn}$]{\includegraphics[width=3.4in]{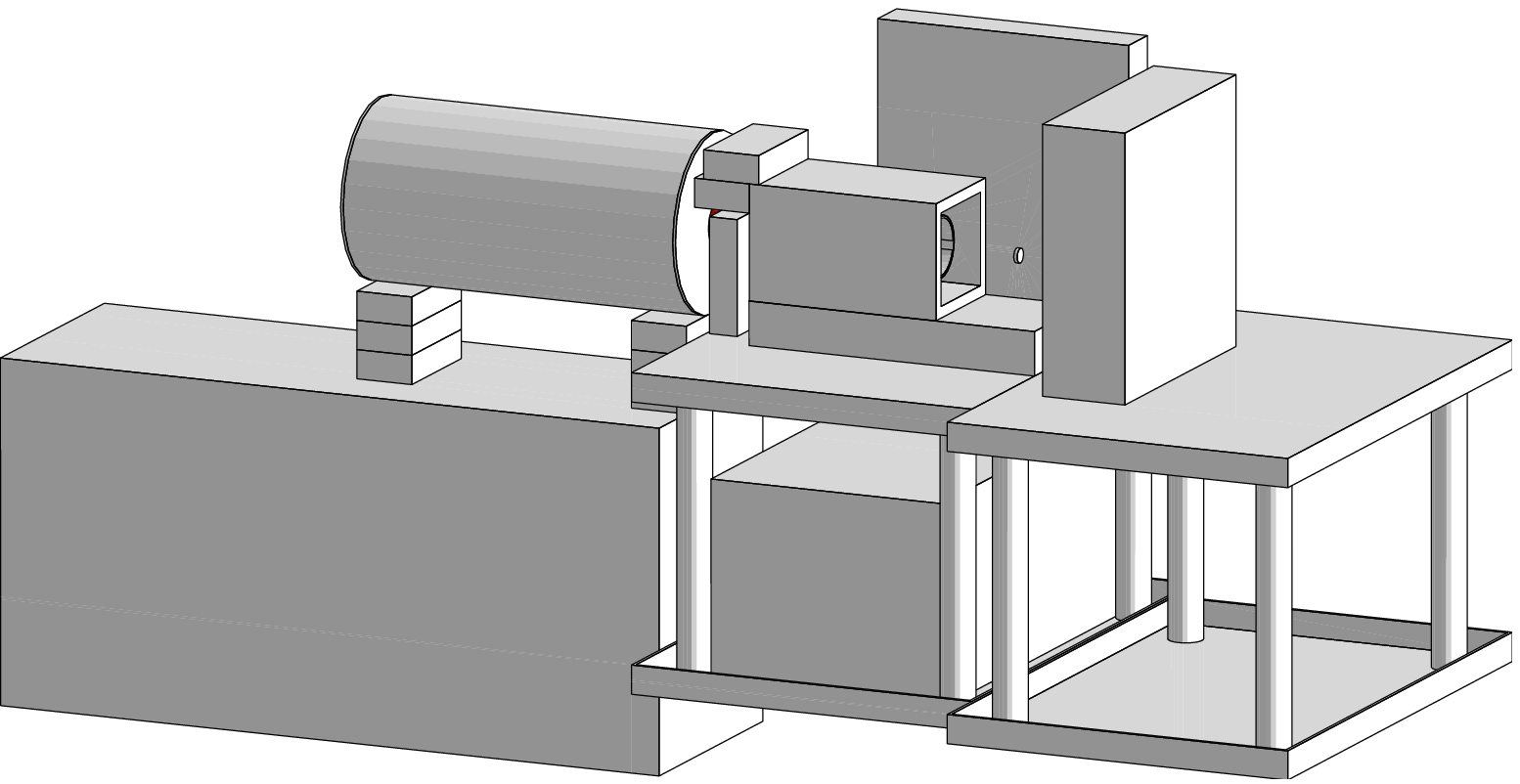}%
\label{fig:sim_setup2}}}
\caption{The geometrical model of the experiment used for the Geant4 simulations. The detector is located inside the hollow lead blocks at the image center, its dewar extends to the left and sits on top of borated wax blocks, which in turn are placed upon a wooden crate. The shielding is placed on top of two steel tables.}
\label{fig:sim_setup}
\end{figure*}

\section{Simulation of the Experiment}

The validation process involves the comparison of simulated and experimental spectra. A Geant4-based application was developed for this purpose, which models the experimental set-up and produces simulated spectra in the same format as experimental ones.
The results reported in this paper were produced in a simulation environment based on Geant4, version 9.4p04. The simulations were executed on a 16-core XEON machine. 

\subsection{Strategy}
Three distinct simulation scenarios are involved in the radioactive decay validation process: with per-decay sampling as implemented in Geant4 original RDM package, with per-decay sampling as implemented in the RDM-extended package and with statistical sampling, which is only available in the RDM-extended package. It was verified\cite{RadDecay2012_1} that the original and refactored implementations of per-decay sampling produce identical results, as expected from a refactoring process~\cite{fowler1999refactoring}, which is meant to improve the design of existing software without modifying its functionality. Therefore this paper reports validation results related to two radioactive decay simulation configurations only: with statistical sampling, which are pertinent to the RDM-extended code only, and with per-decay sampling, which are pertinent to both the original RDM and RDM-extended packages. 

All experimental test cases were simulated using both per-decay sampling and statistical sampling. Simulations with statistical sampling were performed with the RDM-extended code. Simulations using per-decay sampling of radioactive decays were performed with the RDM code in the released Geant4 9.4p04 release. The reported validation results also apply to the implementation in Geant4 versions 9.5 and 9.6 (the latest at the time of writing this paper), which, aside from the inclusion of the handling of forbidden $\mathrm{\beta}$-decays, use the same radioactive decay simulation algorithms and associated data as in Geant4 9.4p04.

Using the geometry described in Section~\ref{sec:sim_geom}, each calibration source decayed a simulated $5\times 10^{6}$ times and the energy deposited in the Ge-crystal by the resulting decay radiation was recorded. These energy depositions were binned into spectra, which were then processed to include the detector response as described in Section~\ref{sec:sim_response}. For the analysis and response adjustment custom IDL (Interactive Data Language) programs, as well as the HYPERMET program were used.

\subsection{Geometrical Model}
\label{sec:sim_geom}
Comparisons of $\mathrm{\gamma}$-ray detectors with Geant4 simulations in previous works, e.g.~\cite{Hurtado5076086, 2004NIMPA.518..764H, Laborie2002618, Golovko2008266, bissaldi_rad}, have shown that the simulation outcome can be sensitive even to small changes in the geometric representation of the detector. Accordingly, care was taken to model the detector, its shielding and the other objects in proximity, such as the tables, accurately and in detail as shown in Fig.~\ref{fig:sim_setup}. Measurements were either taken directly from the experimental setup, or in case of the detector head obtained from the manufacturer.

In contrast to previous works no calibration of the simulation to the measurements was performed. This is usually done by iteratively altering the mass model within reasonable error bounds. Specifically, this means that none of the dimensions taken from the experimental setup were altered. Only the thickness of the active deposit of the source was allowed to vary by as much as $50\%$, which improved the simulation's capability to reproduce the X-ray fluorescence lines of $\mathrm{^{133}Ba}$ and $\mathrm{^{137}Ca}$ which are located below $40\,\mathrm{keV}$. This alteration is to account for uncertainties in the active deposit thickness, which could not be measured directly due to the fact that it is encapsulated. These variations can be justified as it is not the simulated detector itself which is modified.

A point of concern was how accurate the model of the lab room would need to be. Other authors, e.g. Bissaldi~{\it et al.}~\cite{bissaldi_rad}, have found that $\mathrm{\gamma}$-rays that are scattered from the ceiling, the floor or the lab walls can contribute in the order of $10\%$ to the total radiation registered by the simulated detector. Comparisons of simulations that neglected or included the lab room environment in the geometry model showed that this contribution was less than $3\%$ in our case. This lower value can be attributed to the presence of a collimator and additional shielding surrounding the source. Consequently, the room was modeled in a simplified fashion, consisting only of perpendicular arrangements of concrete walls, floor and ceiling. Laboratory equipment other than the detector and the tables and crates it was placed upon was not included in the simulation.

\subsection{Physics Configuration}
Except for the two distinct radioactive decay codes, both sets of simulations used an identical physics setup using the {\it G4EmLivermorePhysics} physics list included in Geant4 9.4p04 to simulate electromagnetic physics. Auger-electron and fluorescence production was activated using the respective physics options. Hadronic physics processes were not used, as hadron production, i.e. $\mathrm{\alpha}$-emission, does not occur for the measured isotopes. The default cut value was set to $10\,\mathrm{\mu m}$; step limits were not imposed.

\subsection{Detector Response and Simulation Analysis}

\label{sec:sim_response}
The data output by the Geant4 simulations consisted of a list of energy-depositing events registered in the detector and includes information such as event number, deposited energy and particle type. For the actual comparison this event list was binned into a spectrum and the detector's response characteristics were taken into account. For the binning the $8192$ channel scheme and corresponding energy calibration of the experiment were adopted. The Geant4 simulations do not take into account all characteristics of the detector's response function, as described in Section~\ref{sec:exp_analysis}. Accordingly, the simulated, binned spectrum needs to be re-processed to include features such as peak broadening. The parameters necessary for these computations were obtained from the measurements. They are shown in Table~\ref{tab:parameters_1} as well as Fig.~\ref{fig:fwhm_fit},\ref{fig:alpha_fit} and~\ref{fig:beta_fit}.

In contrast to the aforementioned peak characteristics, the Compton-continuum is simulated by Geant4 directly. Here only the addition of exponential components from the peaks and a broadening due to electronic noise needs to be considered as part of data preparation. It is worthwhile to note that all the above processing steps were implemented in a fashion that conserves the sum of registered simulated counts. 

The reprocessed spectra were then analyzed using the HYPERMET program in the same fashion the experimental data have been analyzed, accordingly yielding a similar set of parameters which could be used for further comparisons.

\subsection{Epistemic Uncertainties}

For the isotopes measured as part of this work the verification results from~\cite{RadDecay2012_1} are summarized in Table~\ref{tab:ensdf_comp}. Above $50\,\mathrm{keV}$ the radiation intensity deviations both codes produce are below $5\%$; for the RDM-extended this is also true at lower (X-ray) energies. Accordingly, deviations between simulation and measurement of the same order of magnitude are to be expected. Additional application-dependent offsets between simulations and experiment may be introduced by the following:

\begin{itemize}
 \item Inherent uncertainties in the various other processes simulated by Geant4, such as Compton-scattering, pair-production and the photo-electric effect, especially. The scattering processes are responsible for the shape of the continuum, as they remove events from the photo peaks.
 \item Small geometrical details which cannot be inferred from measurement, for example the dead layer of the Germanium-crystal or the thickness of the active deposit in the radioactive source. Uncertainties here will manifest themselves at lower energies.
% \item  The lab room, which was simplified for the simulation model, will introduce backscattered events onto the detector. 
 \item  Inaccurate modeling of the experimental detector's response function. Geant4 does not simulate semi-conductor physics nor are the effects of the read-out electronics included in the Monte-Carlo simulation. Thus the true response must be inferred from measurements and modelled as part of the simulation analysis. An error here would manifest itself for instance in an energy dependent offset of the detector's energy resolution.
 \item  Knowledge about the absolute intensity of the measured isotopes which is needed to normalize the simulation to the experiment. The date of the reference measurements is known with a precision of one month with an additional error of up to $5\%$ for the magnitude of the activity. This systematic error would be observed by a constant offset of photo peaks and continuum when comparing simulation and measurement.
\end{itemize}

\section{Results}
\label{sec:results}
The physics associated with radioactive decay determines the presence of characteristic structures (photo-peaks) in the measured spectrum, and their features: the purpose of the validation process consists of determining the compatibility between the physics model of radioactive decay in the simulation and the features of the measurements that are intrinsic to the physics of radioactive decay. 

Other features of the observed spectrum (the ''continuum'') are due to physics other than radioactive decay (namely interactions of photons and secondary electrons with matter) and instrumental effects (detector response, efficiency etc.). This other physics is not the object of validation in this paper; validation of Geant4 electromagnetic physics is documented in~\cite{batic2012photon,pia2011evaluation,lechner2009validation,pia2009validation,guatelli2007geant4,Pia4237413}. 

Section V.A deals with the analysis of photo-peaks. Compatibility between the measured and simulated continuum spectrum, discussed in section V.B, shows that the simulation as a whole (geometry, physics settings other than radioactive decay, detector response etc.) adequately describes the experiment; these results support the correctness of the methodology on which the validation of Geant4 radioactive decay modeling is based.

\begin{figure*}
\centerline{\subfloat{\begin{tikzpicture}%
\node[above right] (img) at (0,0) {\includegraphics[width=3.4in]{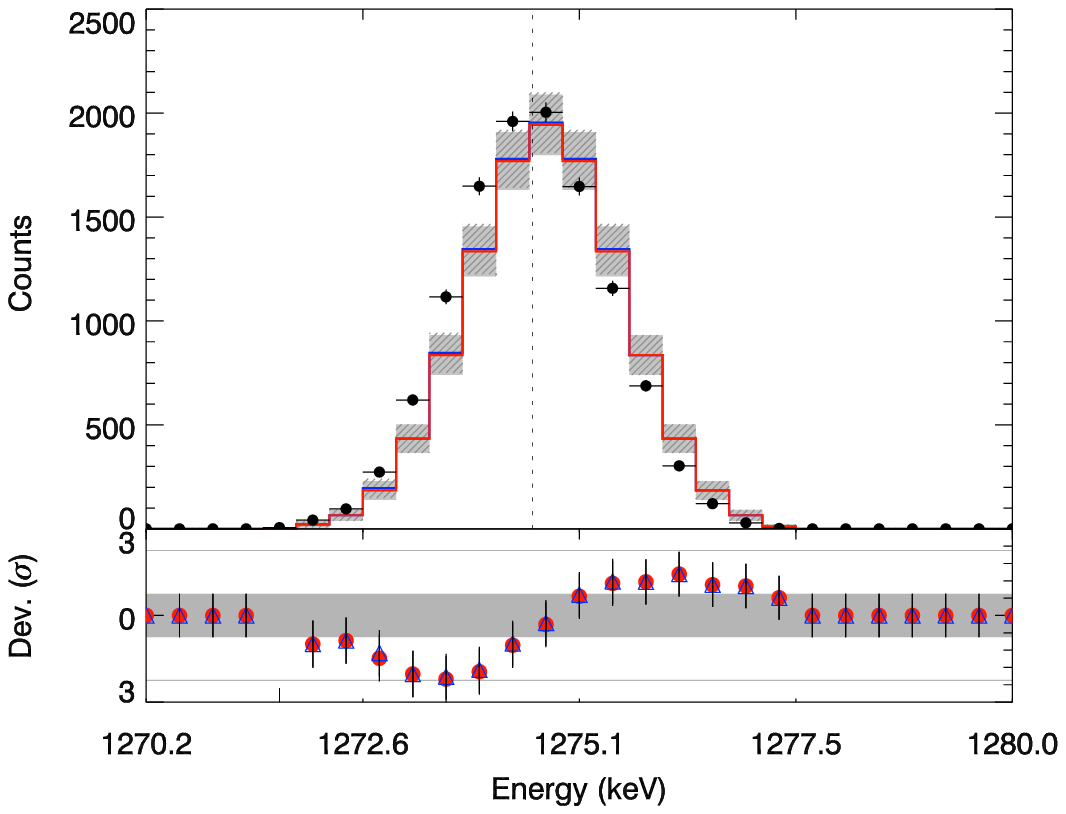}};%
\node at (70pt,170pt) {$\mathrm{^{22}Na}$};%
\end{tikzpicture}}
\hfil
\subfloat{\begin{tikzpicture}%
\node[above right] (img) at (0,0) {\includegraphics[width=3.4in]{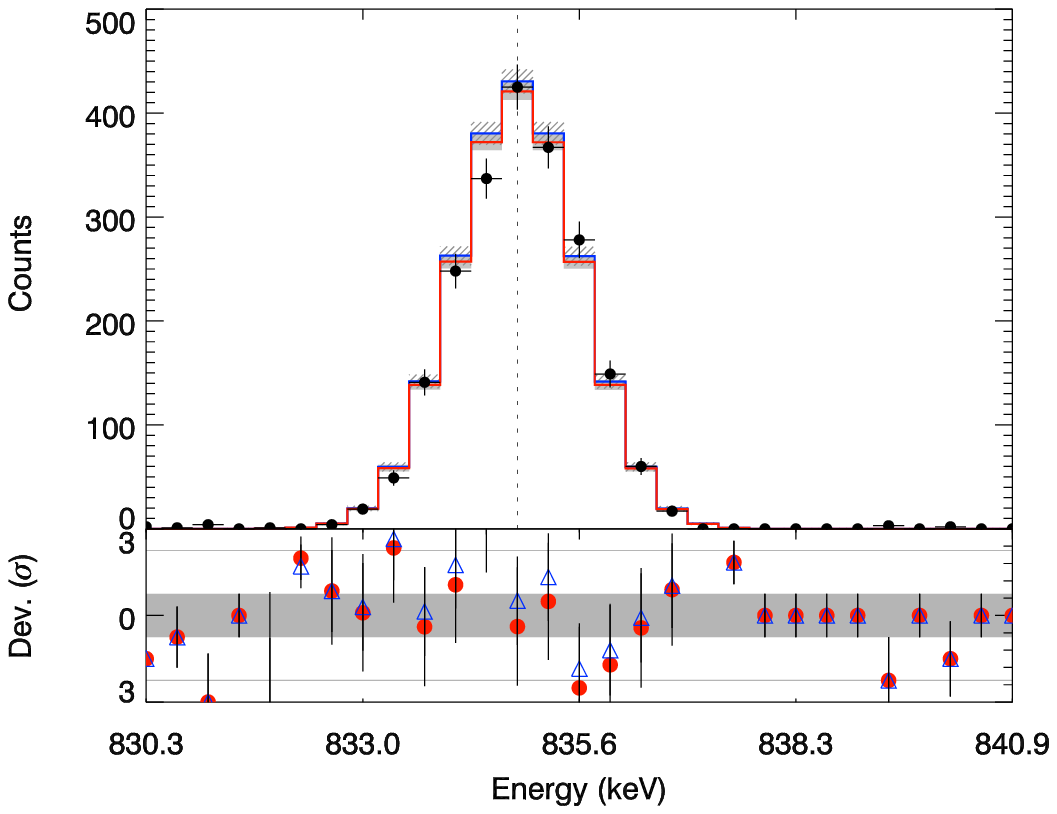}};%
\node at (70pt,170pt) {$\mathrm{^{54}Mn}$};%
\end{tikzpicture}}%
}
\centerline{\subfloat{\begin{tikzpicture}%
\node[above right] (img) at (0,0) {\includegraphics[width=3.4in]{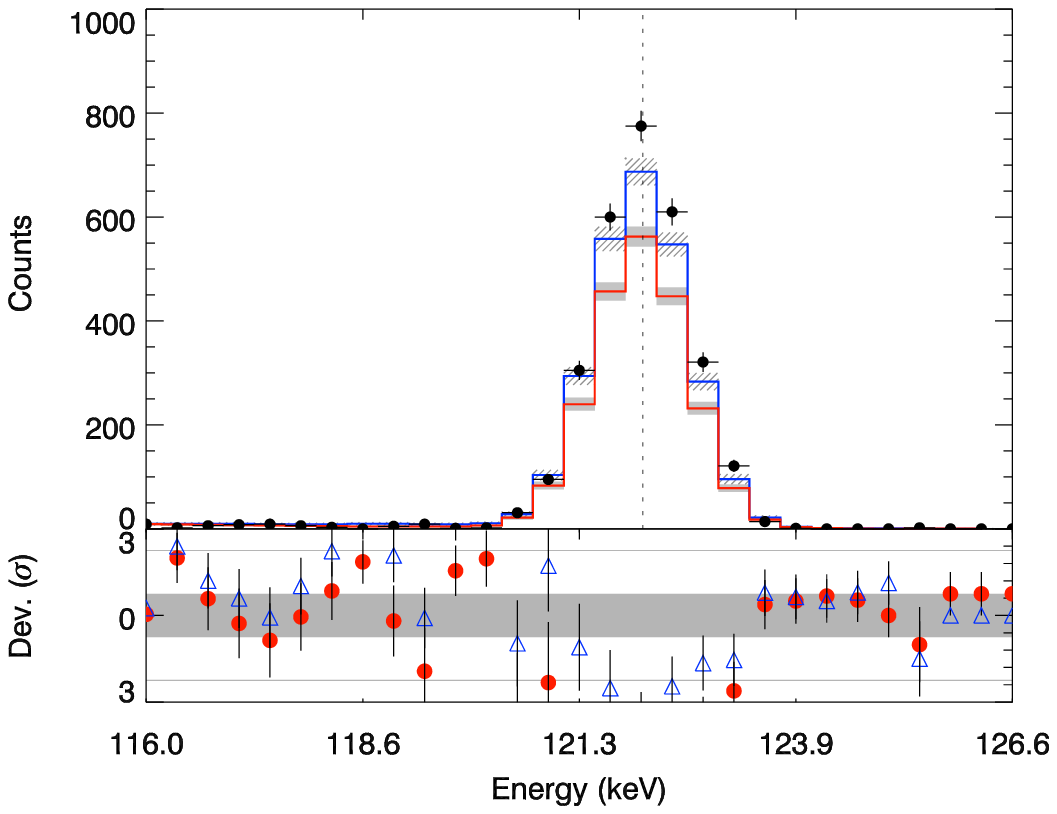}};%
\node at (70pt,170pt) {$\mathrm{^{57}Co}$};%
\end{tikzpicture}}
\hfil
\subfloat{\begin{tikzpicture}%
\node[above right] (img) at (0,0) {\includegraphics[width=3.4in]{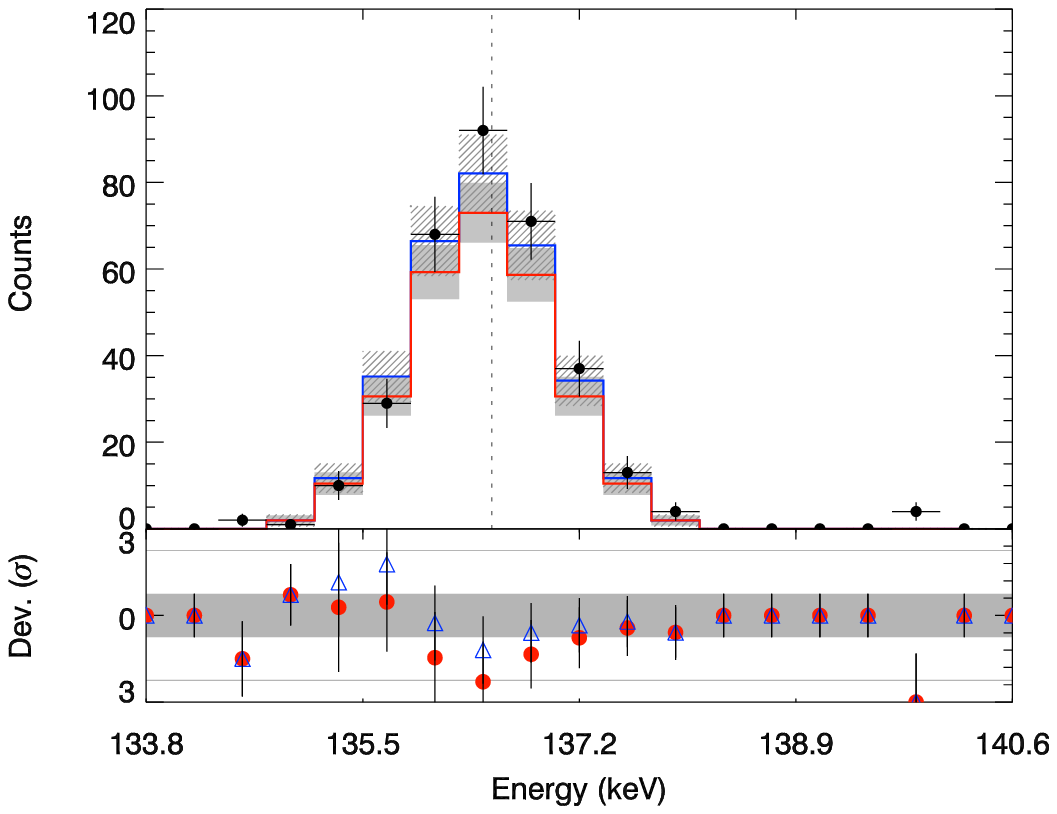}};%
\node at (70pt,170pt) {$\mathrm{^{57}Co}$};%
\end{tikzpicture}}%
}
\centerline{\subfloat{\begin{tikzpicture}%
\node[above right] (img) at (0,0) {\includegraphics[width=3.4in]{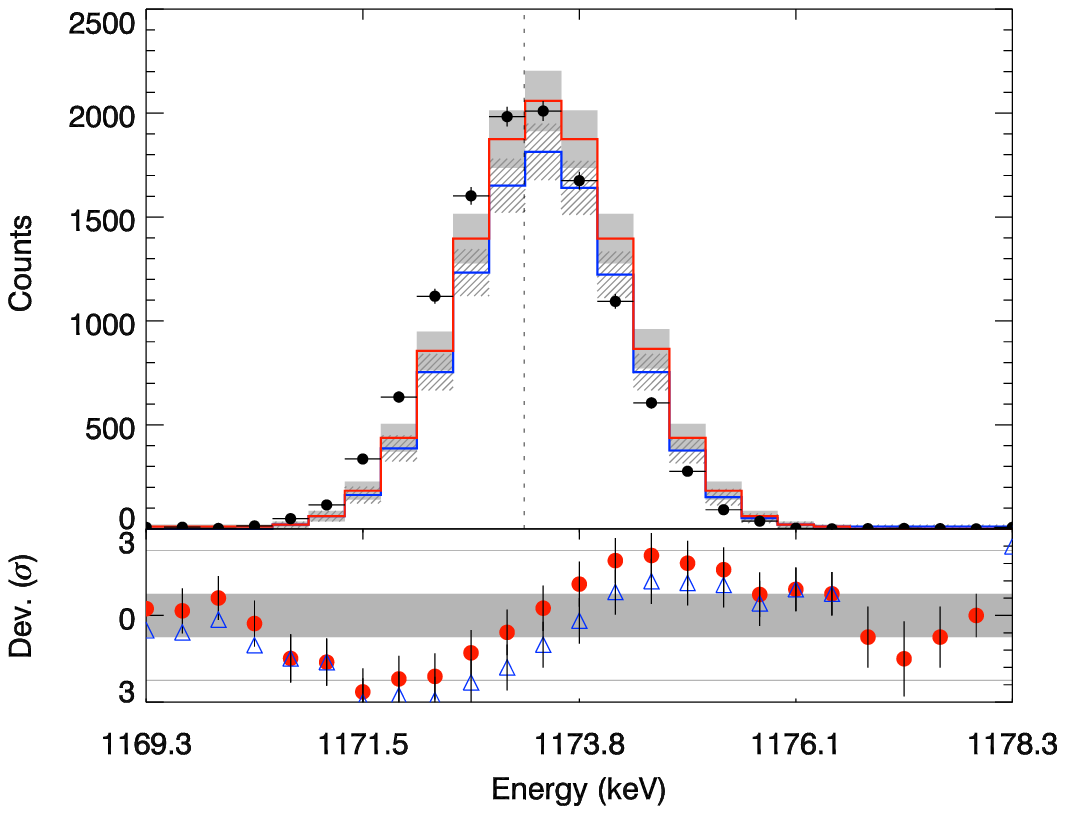}};%
\node at (70pt,170pt) {$\mathrm{^{60}Co}$};%
\end{tikzpicture}}
\hfil
\subfloat{\begin{tikzpicture}%
\node[above right] (img) at (0,0) {\includegraphics[width=3.4in]{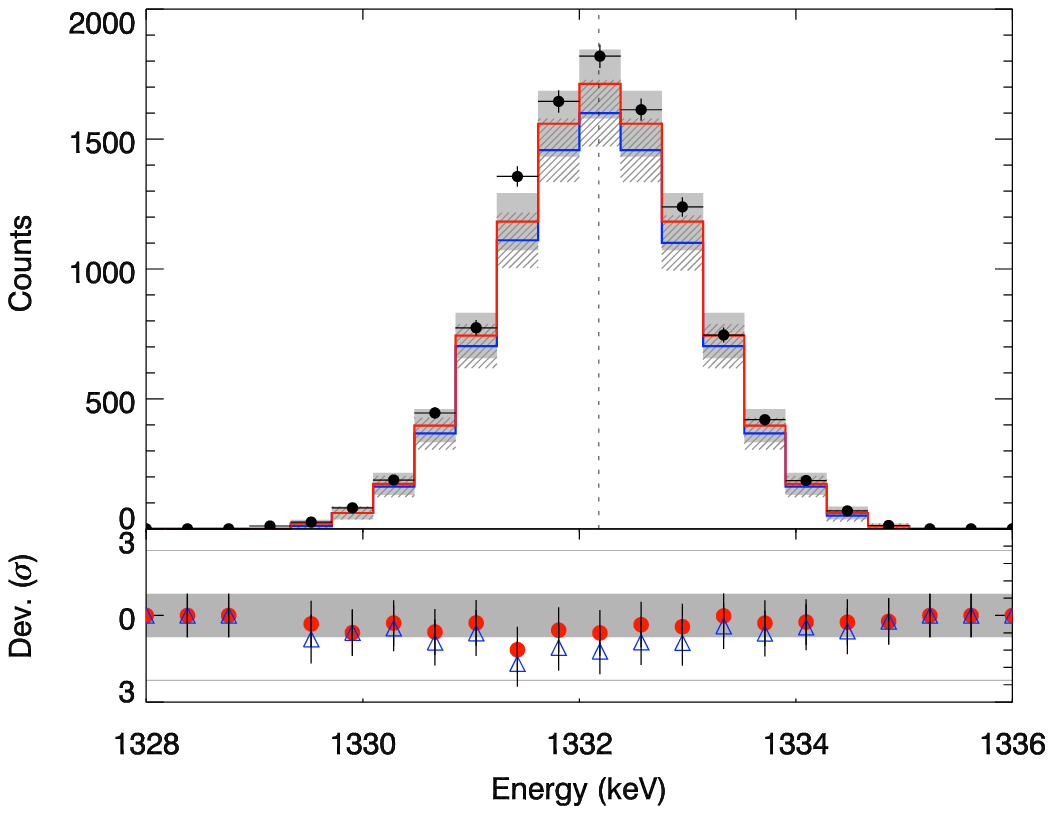}};%
\node at (70pt,170pt) {$\mathrm{^{60}Co}$};%
\end{tikzpicture}}%
}
\caption{Comparison of the simulated and measured photo peak regions. From top left to bottom right the following isotopes and peaks are shown: $\mathrm{^{22}Na_{1274.54\,keV}}$, $\mathrm{^{54}Mn_{834.85\,keV}}$, $\mathrm{^{57}Co_{122.06\,keV}}$, $\mathrm{^{57}Co_{136.47\,keV}}$, $\mathrm{^{60}Co_{1173.23\,keV}}$, $\mathrm{^{60}Co_{1332.49\,keV}}$. Simulations using the per-decay sampling (red line, hatched $1\,\mathrm{\sigma}$ errors) and the RDM-extended with statistical sampling (blue line, filled $1\,\mathrm{\sigma}$ errors) are compared to experimental data. The lower panel shows the residuals (blue triangles: per-decay sampling, filled read circles: RDM-extended; not displayed data--points lie outside the shown deviation range) in terms of $\mathrm{\sigma}$ uncertainties (filled area: $1\mathrm{\sigma}$, horizontal lines: $3\mathrm{\sigma}$). The vertical dashed lines show the peak positions found in the literature.}
\label{fig:peakcomp1}
\end{figure*}

\begin{figure*}
\centerline{\subfloat{\begin{tikzpicture}%
\node[above right] (img) at (0,0) {\includegraphics[width=3.4in]{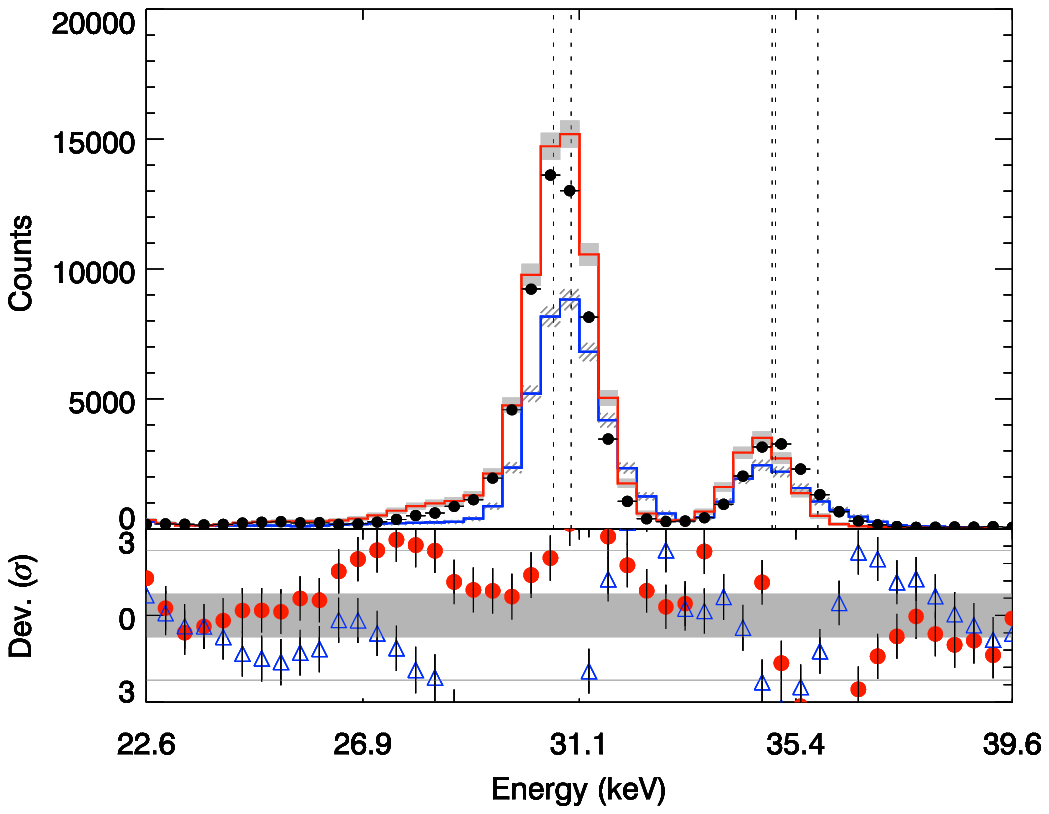}};%
\node at (70pt,170pt) {$\mathrm{^{133}Ba}$};%
\end{tikzpicture}}
\hfil
\subfloat{\begin{tikzpicture}%
\node[above right] (img) at (0,0) {\includegraphics[width=3.4in]{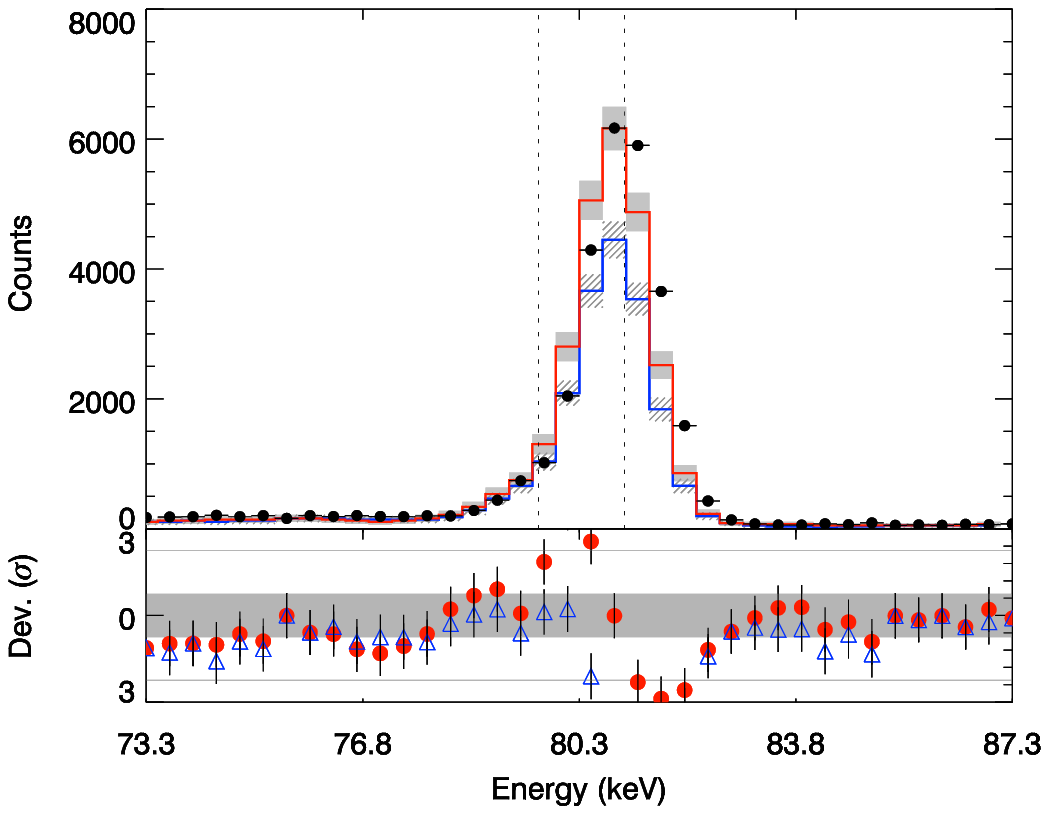}};%
\node at (70pt,170pt) {$\mathrm{^{133}Ba}$};%
\end{tikzpicture}}%
}
\centerline{\subfloat{\begin{tikzpicture}%
\node[above right] (img) at (0,0) {\includegraphics[width=3.4in]{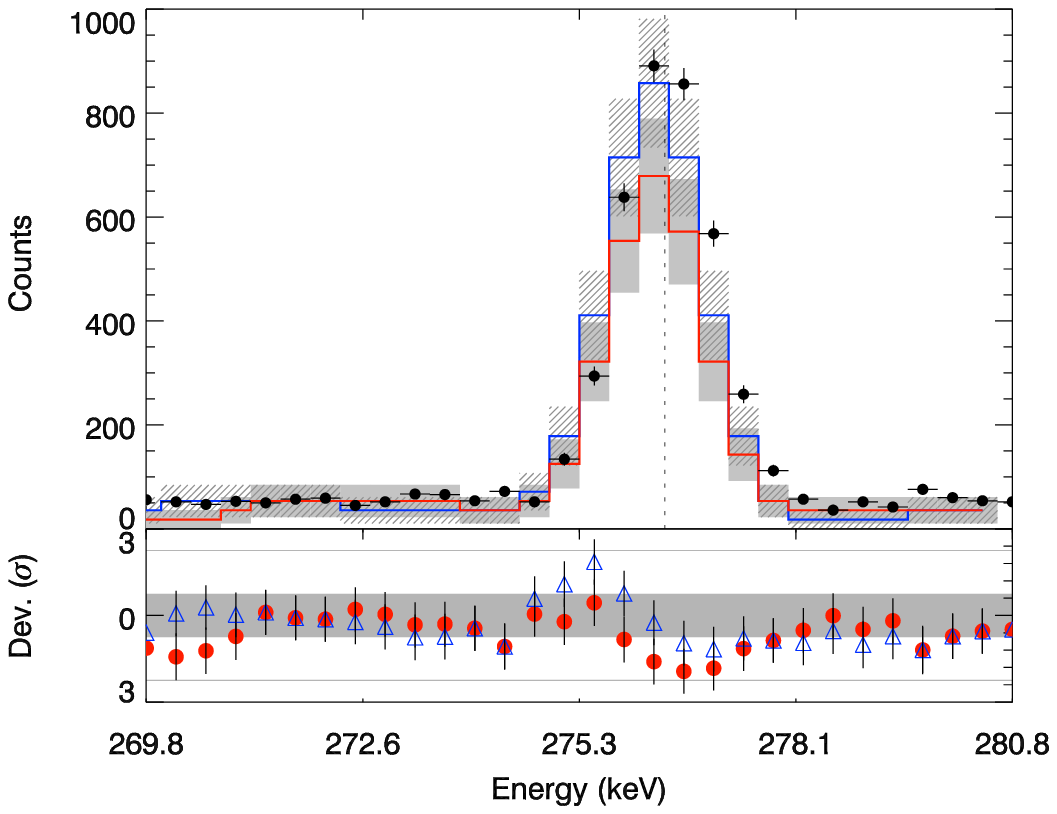}};%
\node at (70pt,170pt) {$\mathrm{^{133}Ba}$};%
\end{tikzpicture}}
\hfil
\subfloat{\begin{tikzpicture}%
\node[above right] (img) at (0,0) {\includegraphics[width=3.4in]{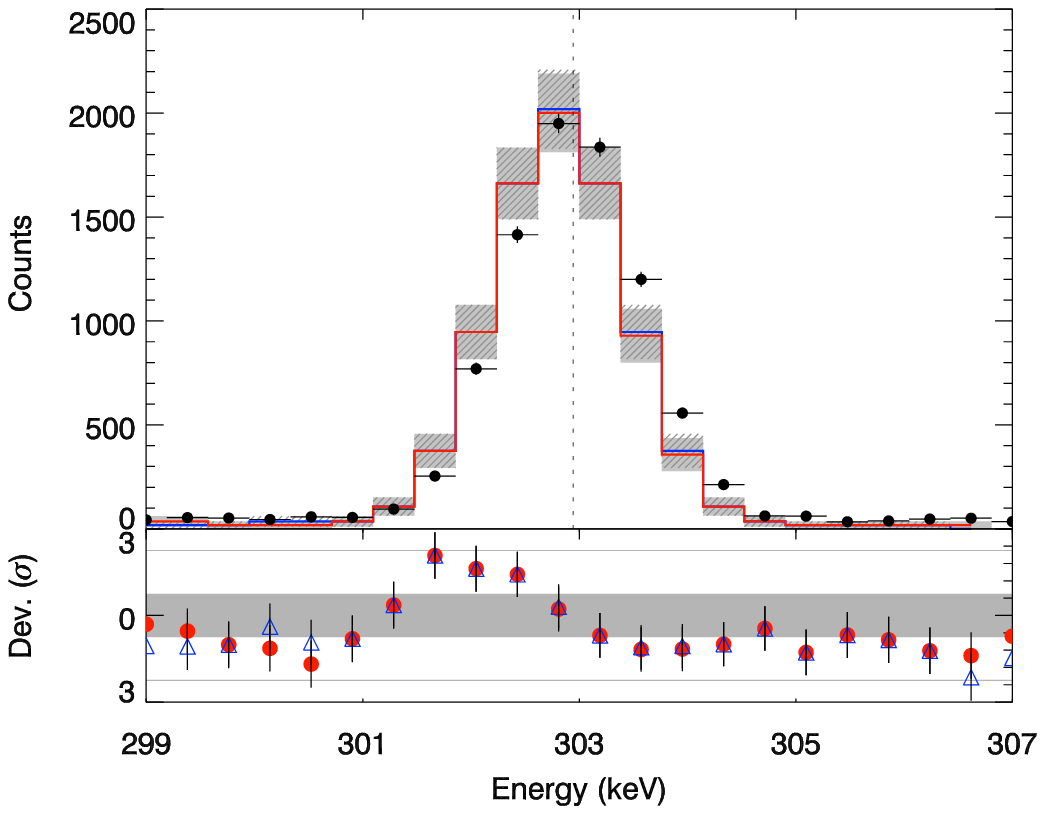}};%
\node at (70pt,170pt) {$\mathrm{^{133}Ba}$};%
\end{tikzpicture}}%
}
\centerline{\subfloat{\begin{tikzpicture}%
\node[above right] (img) at (0,0) {\includegraphics[width=3.4in]{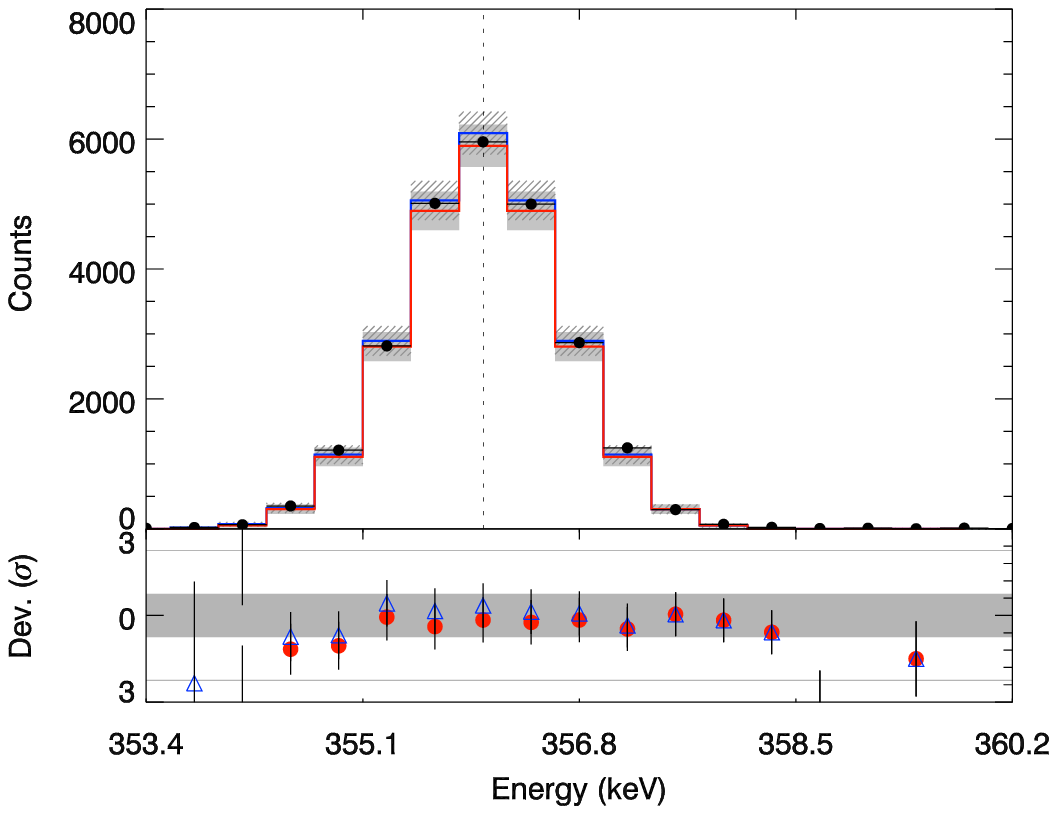}};%
\node at (70pt,170pt) {$\mathrm{^{133}Ba}$};%
\end{tikzpicture}}
\hfil
\subfloat{\begin{tikzpicture}%
\node[above right] (img) at (0,0) {\includegraphics[width=3.4in]{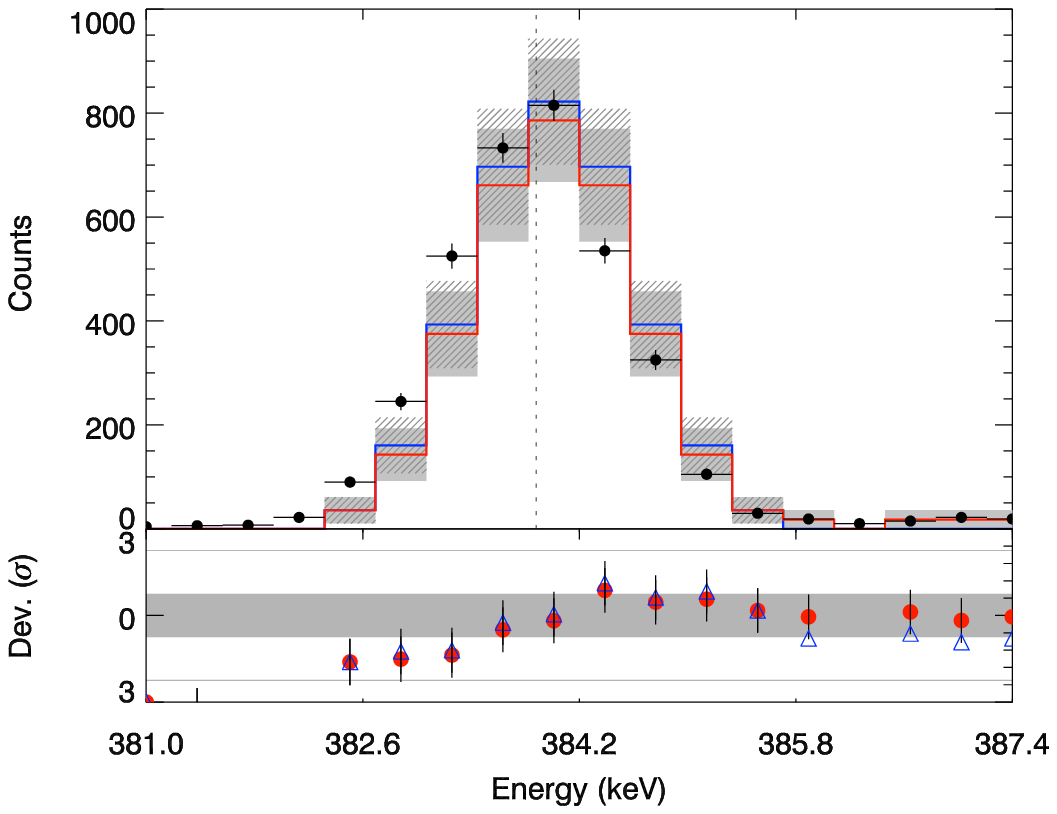}};%
\node at (70pt,170pt) {$\mathrm{^{133}Ba}$};%
\end{tikzpicture}}%
}
\caption{Comparison of the simulated and measured photo peak regions. From top left to bottom right the following isotopes and peaks are shown: $\mathrm{^{133}Ba_{30.63-35.82\,keV}}$, $\mathrm{^{133}Ba_{79.61\&80.99\,keV}}$, $\mathrm{^{133}Ba_{276.40\,keV}}$, $\mathrm{^{133}Ba_{302.85\,keV}}$, $\mathrm{^{133}Ba_{356.01\,keV}}$, $\mathrm{^{133}Ba_{383.85\,keV}}$. Simulations using the per-decay sampling (red line, hatched $1\,\mathrm{\sigma}$ errors) and the RDM-extended with statistical sampling (blue line, filled $1\,\mathrm{\sigma}$ errors) are compared to experimental data. The lower panel shows the residuals (blue triangles: per-decay sampling, filled red circles: RDM-extended; not displayed data--points lie outside the shown deviation range) in terms of $\mathrm{\sigma}$ uncertainties (filled area: $1\mathrm{\sigma}$, horizontal lines: $3\mathrm{\sigma}$). The vertical dashed lines show the peak positions found in the literature.}
\label{fig:peakcomp2}
\end{figure*}

\begin{figure*}
\centerline{\subfloat{\begin{tikzpicture}%
\node[above right] (img) at (0,0) {\includegraphics[width=3.4in]{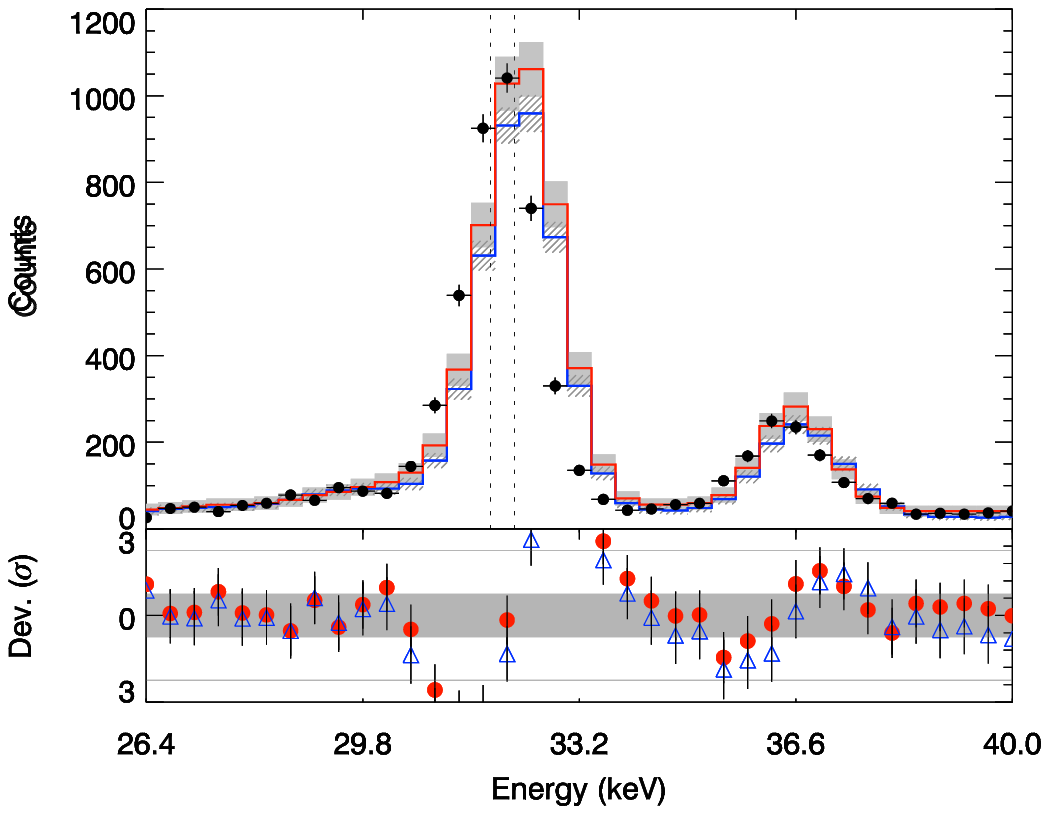}};%
\node at (70pt,170pt) {$\mathrm{^{137}Cs}$};%
\end{tikzpicture}}
\hfil
\subfloat{\begin{tikzpicture}%
\node[above right] (img) at (0,0) {\includegraphics[width=3.4in]{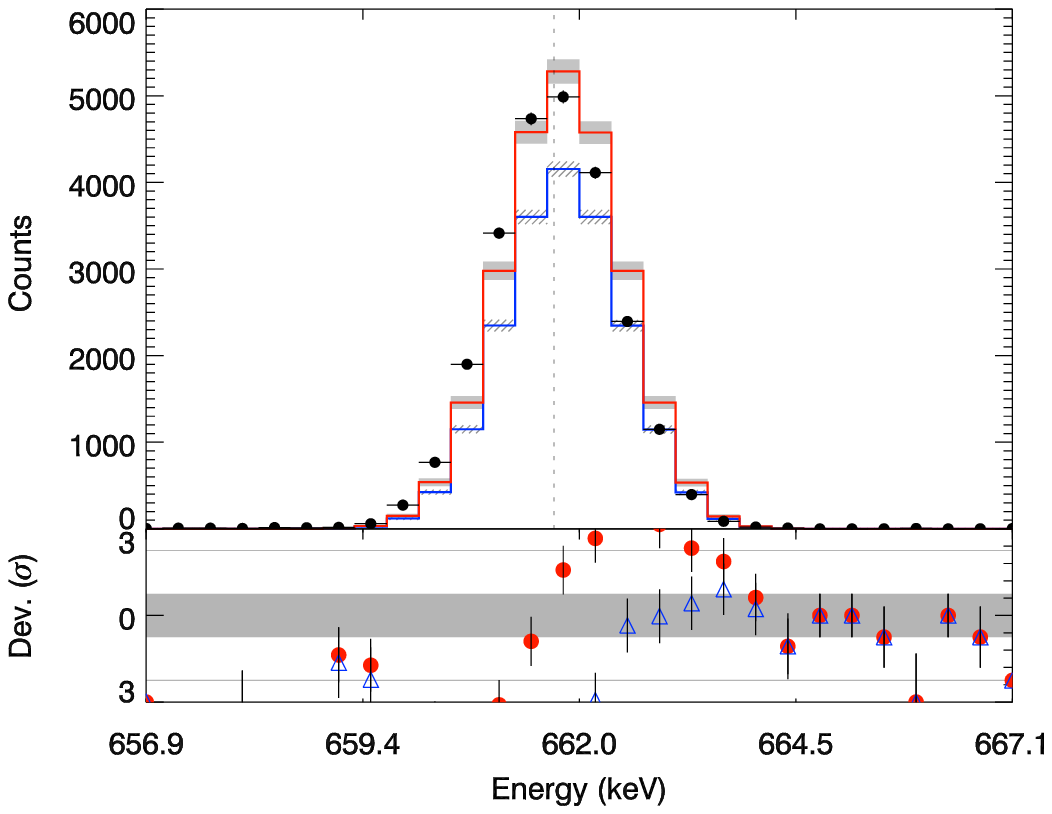}};%
\node at (70pt,170pt) {$\mathrm{^{137}Cs}$};%
\end{tikzpicture}}%
}
\caption{Comparison of the simulated and measured photo peak regions. The following isotopes and peaks are shown: $\mathrm{^{137}Cs_{31.82-37.26\,keV}}$ (left panel), $\mathrm{^{137}Cs_{661.66\,keV}}$ (right panel). Simulations using per-decay sampling (red line, hatched $1\,\mathrm{\sigma}$ errors) and the RDM-extended with statistical sampling (blue line, filled $1\,\mathrm{\sigma}$ errors) are compared to experimental data. The lower panel shows the residuals (blue triangles: per-decay sampling, filled red circles: RDM-extended) in terms of $\mathrm{\sigma}$ uncertainties (filled area: $1\mathrm{\sigma}$, horizontal lines: $3\mathrm{\sigma}$). The vertical dashed lines show the peak positions found in the literature.}
\label{fig:peakcomp3}
\end{figure*}

\subsection{Photo Peaks}
\label{sec:res_peak}
The analysis with the HYPERMET program yields all parameters described in Section~\ref{sec:exp_analysis}, for both for the experimental and the simulated data. The majority of these parameters was used to infer and model the detector response. Accordingly, a comparison of these would yield minor information on the simulation accuracy. Instead we focus on the two quantities which are directly influenced by the Geant4 radioactive decay codes and their data libraries: photo peak intensity (by means of peak area, i.e. the integral over the peak components, i.e. items 1--3 in Section~\ref{sec:exp_analysis}) and peak energy position. Fig.~\ref{fig:peakcomp1}, \ref{fig:peakcomp2} and~\ref{fig:peakcomp3} show a comparison of the photo peak regions between the original RDM per-decay sampling code, the RDM-extended and the measurements. The results are further summarized in Table~\ref{tab:peak_results1}. 

It is obvious from the "S-shape" of some of the residual plots that a peak energy mismatch between simulation and experiment exists. In general a median energy deviation of $(-0.20\pm0.05)\,\mathrm{keV}$ is observed for the per-decay sampling approach and $(-0.18\pm0.05)\,\mathrm{keV}$ for the RDM-extended using statistical sampling. Given that the simulations use evaluated energy positions (indicated as dashed lines in the plots), the deviations likely result from the energy calibration of the measurement, but are well within the energy uncertainties of the binning. Therefore we conclude that both codes model the peak positions equally well.

\begin{figure}
\centering
\centerline{\includegraphics[width=3.4in]{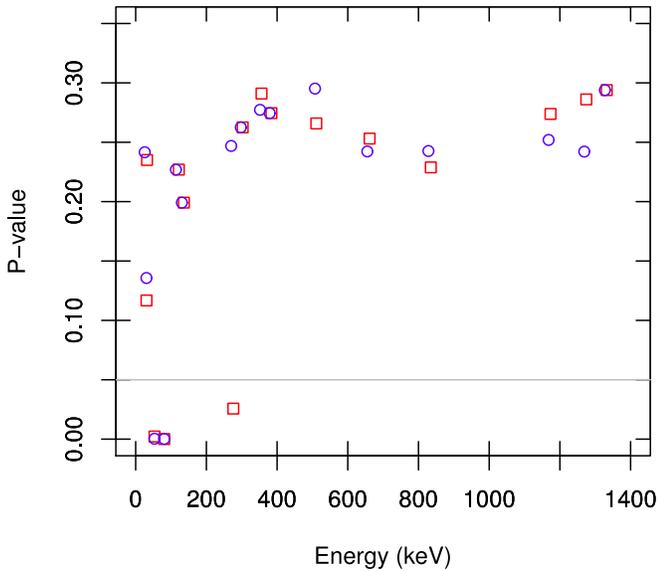}}
\caption{P-values resulting from $\mathrm{\chi^2}$-tests of the individual photopeaks. Shown are the values for the new statistical approach (red squares) and the existing per-decay approach (blue circles). The grey line denotes the $95\mathrm{\%}$ confidence level.}
\label{fig:peak_p_val_chi}
\end{figure}

\begin{figure}
\centering
\centerline{\includegraphics[width=3.4in]{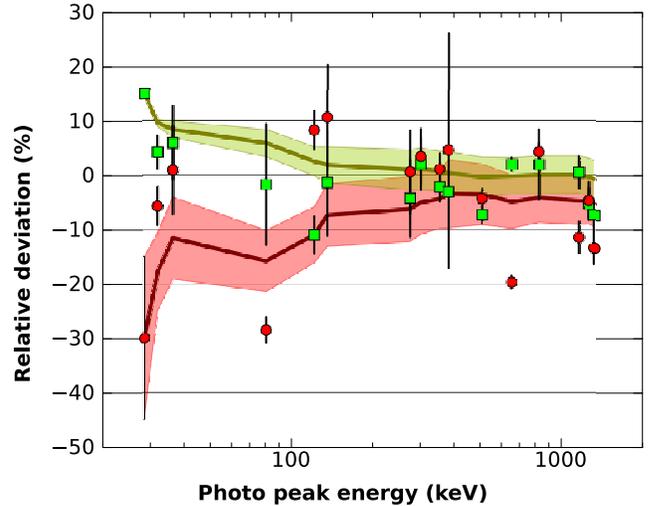}}
%\vspace{-0.5cm}
\caption{The deviations of photo peak areas when comparing per-decay sampling (red) and the RDM-extended using statistical sampling (green) with experimental data. The lines show the running mean deviation, with the corresponding uncertainties shown as shaded areas.}
\label{fig:peak_area_deviations}
\end{figure}

Fig.~\ref{fig:peak_area_deviations} and Table~\ref{tab:peak_results2} give an overview of the photo peak area deviations the original RDM and the RDM-extended code produce with respect to the experiment. These photo peak areas also include the exponential tails described in Section~\ref{sec:exp_analysis}. Together with Fig.~\ref{fig:peakcomp1}, \ref{fig:peakcomp2} and~\ref{fig:peakcomp3}, it is apparent that both codes are capable of reproducing the measured photo peaks within $3\mathrm{\sigma}$ error bounds. The photo peak areas deviate from the measurements by a mean of $(8.65\pm4.56)\%$ when per-decay sampling is used. The RDM-extended's statistical sampling method is capable of modelling the experiment better with a mean deviation of $(4.01\pm3.57)\%$. Accordingly, both codes are generally capable of modeling these photo peaks well within the experimental uncertainties. This reflects itself in the p-values resulting from $\mathrm{\chi^2}$--tests on the data, which are summarized in Fig.~\ref{fig:peak_p_val_chi}. Except for those peaks which have weak intensities or are not separated well from other peaks (resulting in large uncertainties of the response parameters), all p-values are above the $0.05$ significance level. %The aforementioned peaks already exhibited large reduced $\mathrm{\chi^2}$-value when obtaining the model parameters from the experimental data due to their low intensity. 

These semi-qualitative observations are quantitatively and objectively supported by the results of an F-test, which compares the variance of the deviations of experimental and simulated photo-peak areas with per-decay sampling and statistical sampling. The two distributions have been previously verified to have compatible mean values through a two-sample t-test with $0.05$ significance level; a one-sample t--test found that both are compatible with null mean deviation between experimental and simulated photo-peak areas with $0.05$ significance.

The null hypothesis of equivalent variance between the deviations from experiment associated with per-decay and statistical sampling is rejected by the two-tailed F-test with $0.05$ significance: the p-value is $0.011$. A one-tailed F-test rejects the hypothesis of greater variance associated with statistical sampling, in favour of the alternative hypothesis that this algorithm produces smaller variance in the deviation of photo-peak areas from experiment. The significance of the one-tailed F--test is $0.01$ (the resulting p-value is $0.006$).

It is important to note that the results of the F-test do not contradict the results of the $\mathrm{\chi^{2}}$ test reported above, which find the photo-peak areas compatible with experiment with $0.05$ significance for both sampling implementations: the two tests are complementary, as they address different features of the experimental and simulated data samples.

The combination of t--tests and F--test on the photo-peak areas tells us that both sampling algorithms produce deviations from experiment that are in average compatible with zero over the test cases considered in this paper, but the deviations from experiment are more scattered for the per-decay sampling algorithm. Therefore, based on the statistical analysis, one can conclude that the statistical sampling algorithm produces a more accurate simulation of radioactive decay.

Qualitatively, the RDM-extended's statistical sampling performs considerably better at energies below $40\,\mathrm{keV}$. In this X-ray regime, per-decay sampling results in a mean area deviation of $(16.27\pm5.64)\%$. The RDM-extended achieves an accuracy of $(6.73\pm2.52)\%$. In this context it is interesting to note that the deviations produced by per-decay sampling are considerably larger for $\mathrm{^{133}Ba}$ than they are for $\mathrm{^{137}Cs}$. Interestingly these two isotopes are also handled differently in the code's implementation. $\mathrm{^{133}Ba}$ decays via electron capture. In accordance with the code analysis presented in~\cite{RadDecay2012_1} electron capture decays involve delegation of emission production to both the {\it G4Photonevaporation} and {\it G4AtomicDeexcitation} processes; the latter handling the fluorescence emission occurring when the vacated inner electron shell is filled by outer shell electrons. In contrast $\mathrm{^{137}Cs}$ is a $\mathrm{\beta}$-emitter. For $\mathrm{\beta}$-decays the per-decay sampling code delegates deexcitation to the {\it G4Photonevaporation} process. This yields evidence that the {\it G4AtomicDeexcitation} process and its underlying EADL-based data library are at least partially responsible for the large intensity deviation. This observation is in accordance with the findings in~\cite{RadDecay2012_1} and~\cite{pia2011evaluation}.

\begin{table}
\label{tab:peak_results1}
\caption{Energy deviations of the photo peak positions when comparing per-decay sampling and the RDM-extended using statistical sampling with the experimental data.}
\centerline{
\begin{tabular}{lrrr}
&\multicolumn{1}{c}{E (keV)} &\multicolumn{1}{c}{$\mathrm{\Delta E_{old}}$ (keV)} & \multicolumn{1}{c}{$\mathrm{\Delta E_{new}}$ (keV)}\\
\toprule
$\mathrm{^{22}Na}$	& $	1274.74	\pm	0.07	$ & $	0.06	\pm	0.11	$ & $	0.06	\pm	0.11	$\medskip\\
$\mathrm{^{54}Mn}$	& $	834.95	\pm	0.04	$ & $	0.10	\pm	0.06	$ & $	0.10	\pm	0.06	$\medskip\\
$\mathrm{^{57}Co}$	& $	122.05	\pm	0.02	$ & $	-0.36	\pm	0.03	$ & $	-0.36	\pm	0.03	$\\
			& $	136.48	\pm	0.05	$ & $	-0.30	\pm	0.06	$ & $	-0.31	\pm	0.06	$\medskip\\
$\mathrm{^{60}Co}$	& $	1173.41	\pm	0.06	$ & $	-0.01	\pm	0.09	$ & $	-0.02	\pm	0.09	$\\
			& $	1332.73	\pm	0.07	$ & $	0.24	\pm	0.12	$ & $	0.24	\pm	0.12	$\medskip\\
$\mathrm{^{133}Ba}$	& $	30.75	\pm	0.03	$ & $	-0.56	\pm	0.04	$ & $	-0.48	\pm	0.04	$\\
			& $	35.98	\pm	0.08	$ & $	0.66	\pm	0.09	$ & $	0.84	\pm	0.09	$\\
			& $	79.53	\pm	0.05	$ & $	-0.02	\pm	0.08	$ & $	0.03	\pm	0.07	$\\
			& $	81.01	\pm	0.02	$ & $	-0.20	\pm	0.03	$ & $	-0.19	\pm	0.03	$\\
			& $	276.42	\pm	0.02	$ & $	-0.23	\pm	0.02	$ & $	-0.23	\pm	0.03	$\\
			& $	302.85	\pm	0.01	$ & $	-0.26	\pm	0.02	$ & $	-0.24	\pm	0.02	$\\
			& $	356.01	\pm	0.01	$ & $	-0.39	\pm	0.01	$ & $	-0.38	\pm	0.01	$\\
			& $	383.84	\pm	0.02	$ & $	-0.14	\pm	0.02	$ & $	-0.14	\pm	0.03	$\medskip\\
$\mathrm{^{137}Cs}$	& $	32.03	\pm	0.03	$ & $	-0.32	\pm	0.05	$ & $	-0.32	\pm	0.05	$\\
			& $	36.50	\pm	0.05	$ & $	-0.28	\pm	0.07	$ & $	-0.18	\pm	0.06	$\\
			& $	661.69	\pm	0.03	$ & $	-0.07	\pm	0.04	$ & $	-0.07	\pm	0.04	$\\
\bottomrule
\end{tabular}}
\end{table}

\begin{table}
\label{tab:peak_results2}
\caption{Photopeak area deviations when comparing per-decay sampling code and the RDM-extended using statistical sampling with the experimental data.}
\centerline{
\begin{tabular}{lrrr}
& \multicolumn{1}{c}{E (keV)} & \multicolumn{1}{c}{$\mathrm{\Delta I_{old}}$ (\%)} & \multicolumn{1}{c}{$\mathrm{\Delta I_{new}}$ (\%)} \\
\toprule
$\mathrm{^{22}Na}$	& $	1274.74	\pm	0.07	$ & $	-4.70	\pm	3.42	$ & $	-5.35	\pm	2.95	$\medskip\\
$\mathrm{^{54}Mn}$	& $	834.95	\pm	0.04	$ & $	4.26	\pm	1.63	$ & $	1.96	\pm	6.53	$\medskip\\
$\mathrm{^{57}Co}$	& $	122.05	\pm	0.02	$ & $	8.26	\pm	3.60	$ & $	-11.05	\pm	3.44	$ \\
			& $	136.48	\pm	0.05	$ & $	10.67	\pm	9.75	$ & $	-1.40	\pm	9.82	$\medskip\\
$\mathrm{^{60}Co}$	& $	1173.41	\pm	0.06	$ & $	-11.47	\pm	2.93	$ & $	0.47	\pm	2.97	$ \\
			& $	1332.73	\pm	0.07	$ & $	-13.40	\pm	3.02	$ & $	-7.34	\pm	2.96	$\medskip\\
$\mathrm{^{133}Ba}$	& $	30.75	\pm	0.03	$ & $	-29.94	\pm	14.91	$ & $	15.12	\pm	0.17	$ \\
			& $	81.01	\pm	0.02	$ & $	-28.50	\pm	2.36	$ & $	-1.71	\pm	11.17	$ \\
			& $	276.42	\pm	0.02	$ & $	0.57	\pm	7.72	$ & $	-4.24	\pm	7.16	$ \\
			& $	302.85	\pm	0.01	$ & $	3.39	\pm	5.18	$ & $	2.03	\pm	4.81	$ \\
			& $	356.01	\pm	0.01	$ & $	1.09	\pm	2.88	$ & $	-2.23	\pm	2.66	$ \\
			& $	383.84	\pm	0.02	$ & $	4.59	\pm	21.66	$ & $	-3.12	\pm	7.38	$\medskip\\
$\mathrm{^{137}Cs}$	& $	32.03	\pm	0.03	$ & $	-5.72	\pm	3.51	$ & $	4.19	\pm	3.07	$ \\
			& $	36.50	\pm	0.05	$ & $	0.91	\pm	8.16	$ & $	5.91	\pm	6.81	$ \\
			& $	661.69	\pm	0.03	$ & $	-19.69	\pm	1.25	$ & $	1.98	\pm	1.27	$ \\
\bottomrule
\end{tabular}}
\end{table}

\subsection{Continuum}
\label{sec:res_cont}

\begin{figure*}
\centerline{\subfloat{\begin{tikzpicture}%
\node[above right] (img) at (0,0) {\includegraphics[width=3.4in]{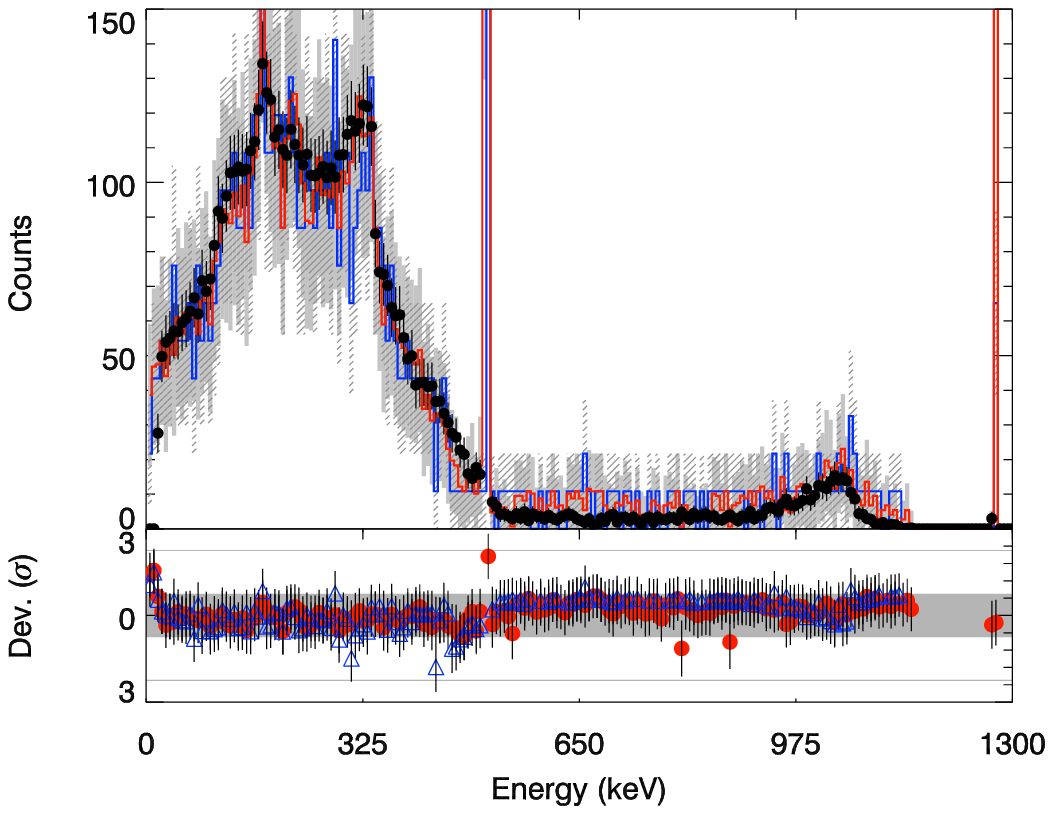}};%
\node at (170pt,170pt) {$\mathrm{^{22}Na}$};%
\end{tikzpicture}}
\hfil
\subfloat{\begin{tikzpicture}%
\node[above right] (img) at (0,0) {\includegraphics[width=3.4in]{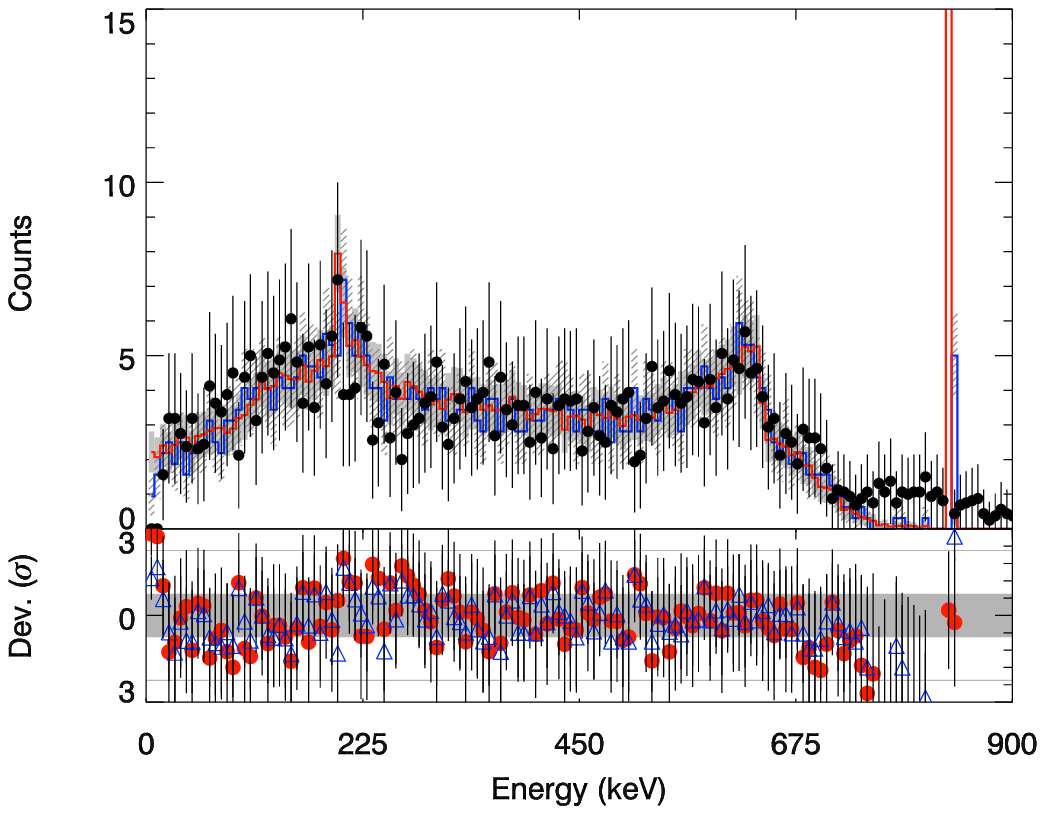}};%
\node at (70pt,170pt) {$\mathrm{^{54}Mn}$};%
\end{tikzpicture}}%
}
\centerline{\subfloat{\begin{tikzpicture}%
\node[above right] (img) at (0,0) {\includegraphics[width=3.4in]{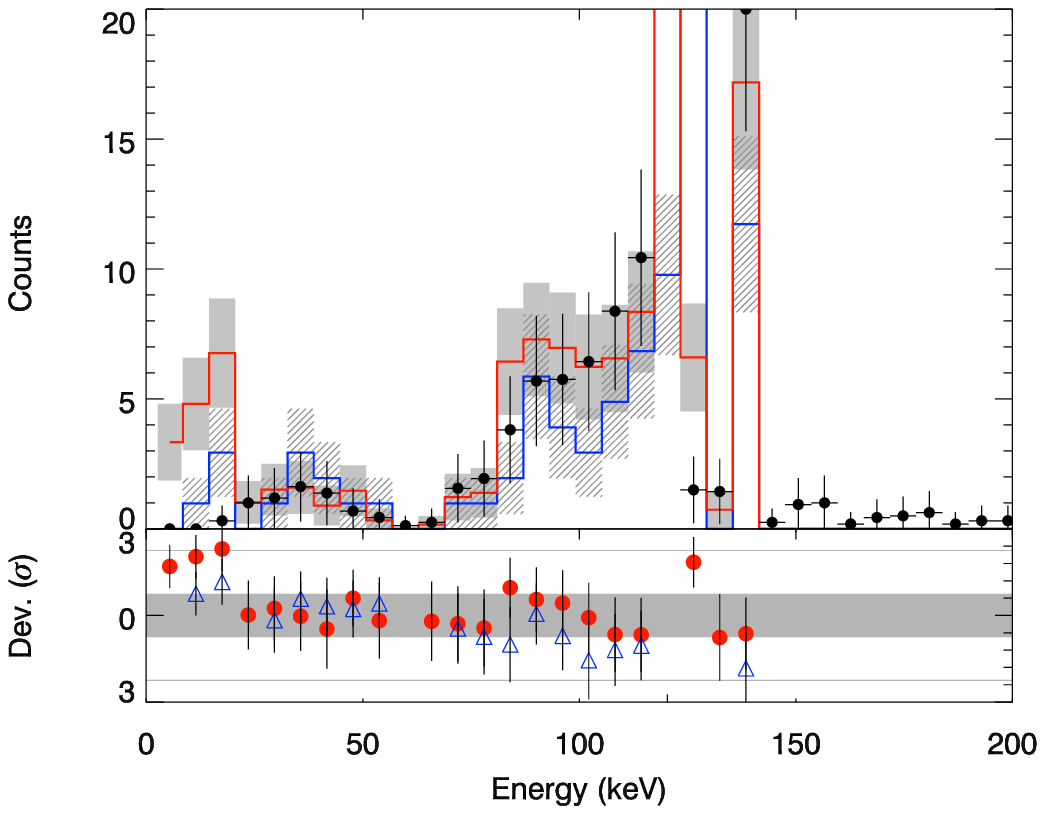}};%
\node at (70pt,170pt) {$\mathrm{^{57}Co}$};%
\end{tikzpicture}}
\hfil
\subfloat{\begin{tikzpicture}%
\node[above right] (img) at (0,0) {\includegraphics[width=3.4in]{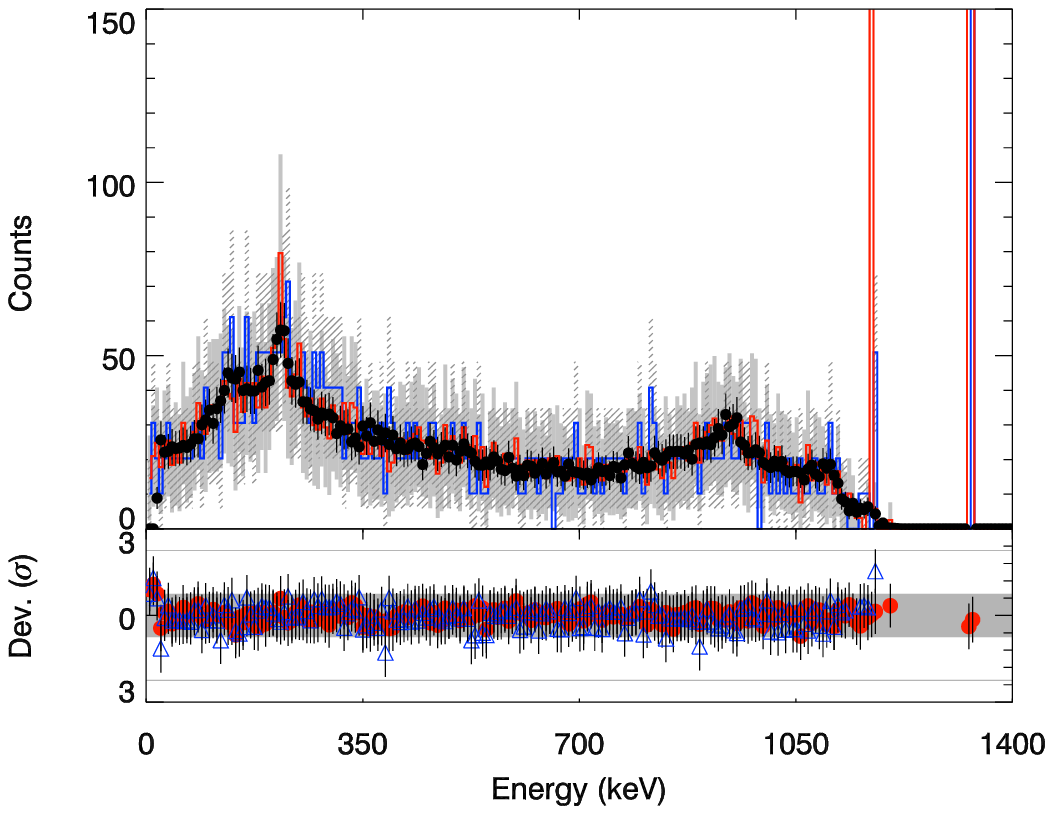}};%
\node at (70pt,170pt) {$\mathrm{^{60}Co}$};%
\end{tikzpicture}}%
}
\centerline{\subfloat{\begin{tikzpicture}%
\node[above right] (img) at (0,0) {\includegraphics[width=3.4in]{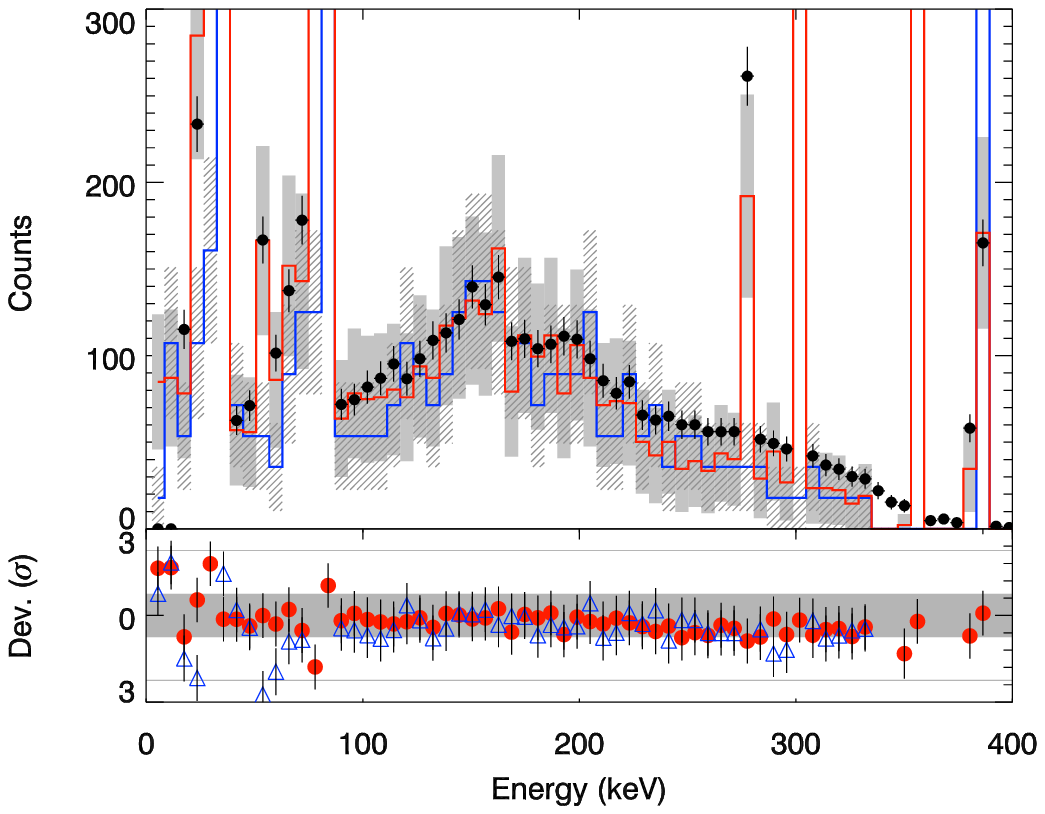}};%
\node at (130pt,170pt) {$\mathrm{^{133}Ba}$};%
\end{tikzpicture}}
\hfil
\subfloat{\begin{tikzpicture}%
\node[above right] (img) at (0,0) {\includegraphics[width=3.4in]{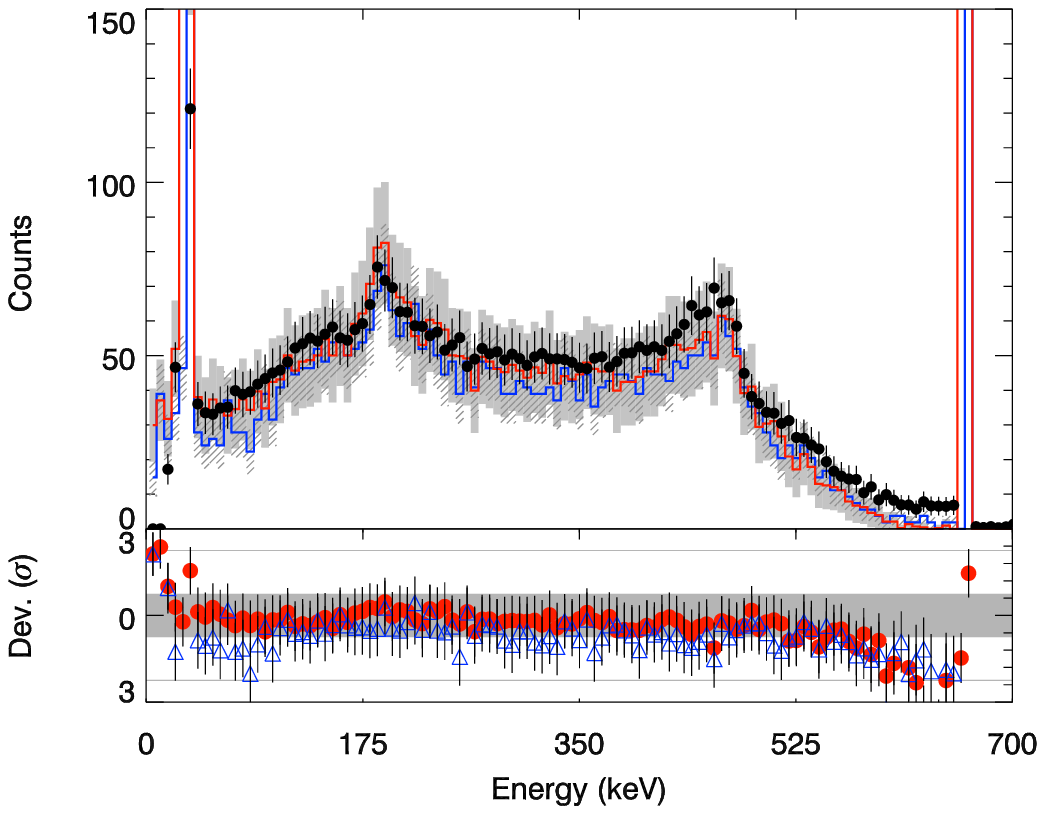}};%
\node at (70pt,170pt) {$\mathrm{^{137}Cs}$};%
\end{tikzpicture}}%
}

\caption{Comparison of the complete simulated and measured spectra. From top left to bottom right the following isotopes and peaks are shown: $\mathrm{^{22}Na}$, $\mathrm{^{54}Mn}$, $\mathrm{^{57}Co}$, $\mathrm{^{60}Co}$, $\mathrm{^{133}Ba}$, $\mathrm{^{137}Cs}$. Simulations using per-decay sampling (red line, hatched $1\,\mathrm{\sigma}$ errors) and the RDM-extended with statistical sampling (blue line, filled $1\,\mathrm{\sigma}$ errors) are compared to experimental data. The lower panel shows the residuals (blue triangles: per-decay sampling, filled red circles: RDM-extended) in terms of $\mathrm{\sigma}$ uncertainties (filled area: $1\mathrm{\sigma}$, horizontal lines: $3\mathrm{\sigma}$).}
\label{fig:contcomp}
\end{figure*}

In contrast to the photo peaks, the inter-peak continuum is less affected by the reprocessing of data to include the full detector response function. Accordingly, this comparison yields more information on Geant4's intrinsic capabilities in modeling the experimental setup. One should note however that the shape of the continuum is much more affected by processes other than the radioactive decay, such as Compton-scattering. Nevertheless its height and also the location of the Compton edges are directly influenced by the photo peak intensity and position sampled by the radioactive decay codes. Accordingly, a comparison of the complete spectra, including the continuum, assesses the performance of the simulation as a whole, with the radioactive decay code playing an integral part as the initial radiation producing process.

\begin{figure}[!h]
\centering
\centerline{\includegraphics[width=3.4in]{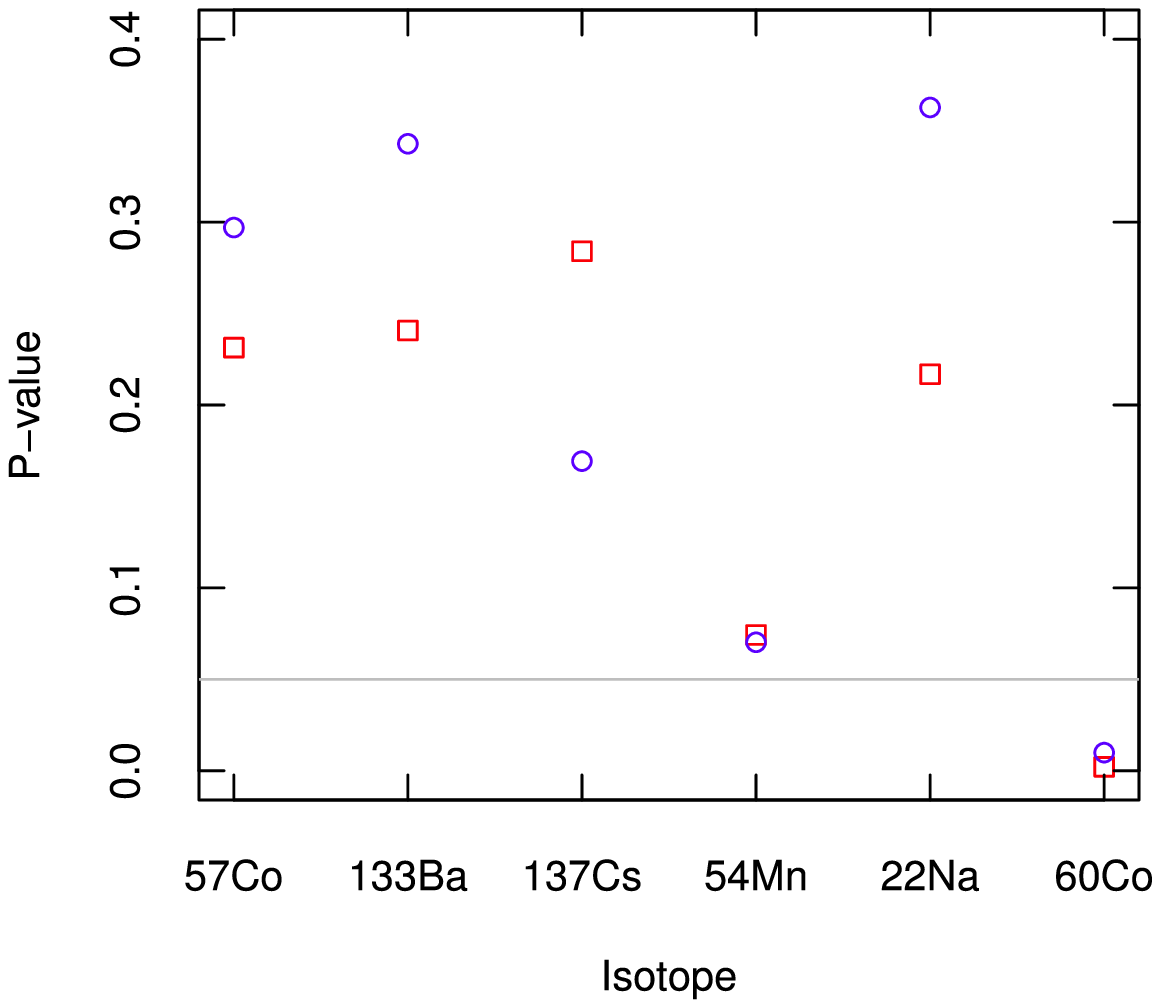}}
\caption{P-values resulting from $\mathrm{\chi^2}$-tests of the continuum. Shown are the values for the new statistical approach (red squares) and the existing per-decay approach (blue circles). The grey line denotes the $95\mathrm{\%}$ confidence level.}
\label{fig:p_val_chi}
\end{figure}

Fig.~\ref{fig:contcomp} shows the complete spectra simulated with the original RDM's per-decay sampling and the RDM-extended statistical sampling in comparison to the measurements. It is apparent from the figure that both codes are capable of reproducing the measurements within $3\mathrm{\sigma}$ uncertainties. This finding is supported by the p-values resulting from comparisons of the continuum using $\mathrm{\chi^2}$--tests, which are shown in Fig.~\ref{fig:p_val_chi}. The continuum is influenced by multiple photo-peaks with varying accuracy, resulting in non-homogeneous offsets throughout the energy-range.% Nevertheless at least one of the two tests is above the $95\mathrm{\%}$ confidence level for each isotope. 

Together with the already discussed peak deviations this leads us to the conclusion that Geant4 is generally capable of reproducing a HPGe-detector measurement of radioactive sources. One should note, however, that especially at lower energies the RDM-extended code using statistical sampling shows significant improvements with respect to the original RDM per-decay sampling implementation.

\section{Conclusion}

In this work a comparison of measurements of radioactive sources with a HPGe-detector with Geant4 simulations of the experimental setup was performed. Two different radioactive decay modeling approaches were used for these simulations: per-decay sampling and statistical sampling. Per-decay sampling is implemented in the original Geant4 RDM package, and has been refactored with equivalent behavior in an extended version of the package, named RDM-extended. Statistical sampling is only available in the RDM-extended package.

Statistical analysis of the experimental and simulated data samples demonstrated that both simulation approaches can reproduce adequately the general features of the measurements.  It also showed that the statistical sampling approach generates a more accurate simulation of the photo-peak areas, i.e. of photo-peak intensities, than the per-decay sampling implementation.

In addition to improved accuracy in the presented validation scenario, the RDM-extended package provides other benefits regarding computational performance and extended functionality. These features are discussed in detail in~\cite{RadDecay2012_1}.

\section*{Acknowledgment}

The authors would like to acknowledge the financial support from the Deutsche Zentrum fuer Luft-- und Raumfahrt (DLR) under Grant number 50QR902 and 50Q1102.

%\ifCLASSOPTIONcaptionsoff
% \newpage
%\fi



% can use a bibliography generated by BibTeX as a.bbl file
\bibliographystyle{IEEEtran}
\bibliography{all}
\end{document}